\documentclass[aps,prb,twocolumn,showpacs,floatfix,groupedaddress,superscriptaddress]{revtex4-2}
\pdfoutput=1
\usepackage[linkcolor=blue,colorlinks=true,breaklinks=true,citecolor=blue,urlcolor=blue]{hyperref}
\usepackage{graphicx,graphics}  
\usepackage{dcolumn}   
\usepackage{braket}
\usepackage[mathscr]{euscript}
\usepackage{amsbsy}
\usepackage{bm}        
\usepackage{amssymb} 
\usepackage{amsmath}
\usepackage{dsfont}
\usepackage{xcolor}
\usepackage{comment}
\usepackage{dirtytalk}
\usepackage[titletoc,title]{appendix}
\usepackage[dotinlabels]{titletoc}
\usepackage[caption=false,labelformat=simple]{subfig}
\begin{document}

%%\preprint{APS/123-QED}

%\title{Topological properties of bilayer \texorpdfstring{$\alpha-\mathcal{T}_3$}{} lattice }% Force line breaks with \\
\title{Topological properties of nearly flat bands in bilayer \texorpdfstring{$\alpha-\mathcal{T}_3$}{} lattice }
\author{Puspita Parui}
    \email{puspitaparui44@gmail.com}
\author{Sovan Ghosh}
    \email{sovanghosh2014@gmail.com}
\author{Bheema Lingam Chittari}%
 \email{bheemalingam@iiserkol.ac.in}
\affiliation{Department of Physical Sciences, Indian Institute of Science Education and Research Kolkata, Mohanpur 741246, West Bengal, India}%



\begin{abstract}
We study the effect of Haldane flux in the bilayer $\alpha$-$\mathcal{T}_3$ lattice system, considering possible non-equivalent, commensurate stacking configurations with a tight-binding formalism. The bilayer $\alpha$-$\mathcal{T}_3$ lattice comprises six sublattices in a unit cell, and its spectrum consists of six bands. In the absence of Haldane flux, threefold band crossings occur at the two Dirac points for both valence and conduction bands. The introduction of Haldane flux in a cyclically stacked bilayer $\alpha$-$\mathcal{T}_3$ lattice system separates all six bands, including two low-energy, corrugated nearly flat bands, and assigns non-zero Chern numbers to each band, rendering the system topological. We demonstrate that the topological evolution can be induced by modifying the hopping strength between sublattices with the scaling parameter $\alpha$ in each layer. In the dice lattice limit ($\alpha = 1$) of the Chern-insulating phase, the Chern numbers of the three pairs of bands, from low energy to higher energies, are $\pm 2$, $\pm 3$, and $\pm 1$. Interestingly, a continuous change in the parameter $\alpha$ triggers a topological phase transition through band crossings between the two lower energy bands. These crossings occur at different values for the conduction and valence bands and depend further on the next nearest neighbor (NNN) hopping strength. At the transition point, the Chern numbers of the two lower conduction and valence bands change discontinuously from $\pm 2$ to $\pm 5$ and $\pm 3$ to $0$, respectively, while leaving the Chern number of the third band intact. 

%\begin{description}
%Use the optional argument of the \verb+\item+ command to give the category of each item. 
%\end{description}
\end{abstract}

%\keywords{Suggested keywords}%Use show keys class option if keyword
\maketitle
%\tableofcontents

\section{\label{sec:level1}Introduction}
The discovery of unconventional superconductivity in twisted-bilayer graphene (TBG) \cite{Cao2018} has ignited significant interest in identifying flat bands in 2D materials \cite{PhysRevMaterials.5.084203}. In magic-angle TBG, flat bands are not isolated unless subjected to a staggered potential and an external electric field \cite{PhysRevB.101.125411}. However, new proposals, such as twisted double bilayer graphene and multilayer graphene aligned on boron nitride, exhibit isolated flat bands under an external electric field \cite{Yankowitz2018, Kim.acs.nanolett.8b03423, PhysRevLett.122.016401, Chen2019, PhysRevB.99.235417, PhysRevB.102.035411, Sinha2020, PhysRevB.103.075423, PhysRevB.103.165112, PhysRevB.104.045413, PhysRevB.105.245124, CHEBROLU2023115526, PhysRevB.108.155406}. These gate-tunable isolated flat bands have been found to possess topological properties \cite{Yankowitz2018, Kim.acs.nanolett.8b03423, PhysRevLett.122.016401, Chen2019, PhysRevB.99.235417, PhysRevB.102.035411, Sinha2020, PhysRevB.103.075423, PhysRevB.103.165112, PhysRevB.104.045413, PhysRevB.105.245124, CHEBROLU2023115526, PhysRevB.108.155406}. The nontrivial flat band topology inherently leads to a rich spectrum of physical phenomena driven by Coulomb interactions, which are proportional to the reduced kinetic energy. Moreover, topological flat bands have enabled the observation of phenomena such as superconductivity \cite{MIYAHARA20071145}, excitonic insulator states \cite{PhysRev.158.462, PhysRevLett.126.196403}, fractional quantum anomalous Hall effects \cite{PhysRevLett.106.236802, PhysRevLett.106.236804, PhysRevLett.106.236803}, ferromagnetism \cite{Stoner.rspa.1938.0066, Jiang2019}, and excited quantum anomalous or spin Hall effects \cite{Zhou_2022}.
\begin{figure*}
    \centering
    \includegraphics[scale = 0.46]{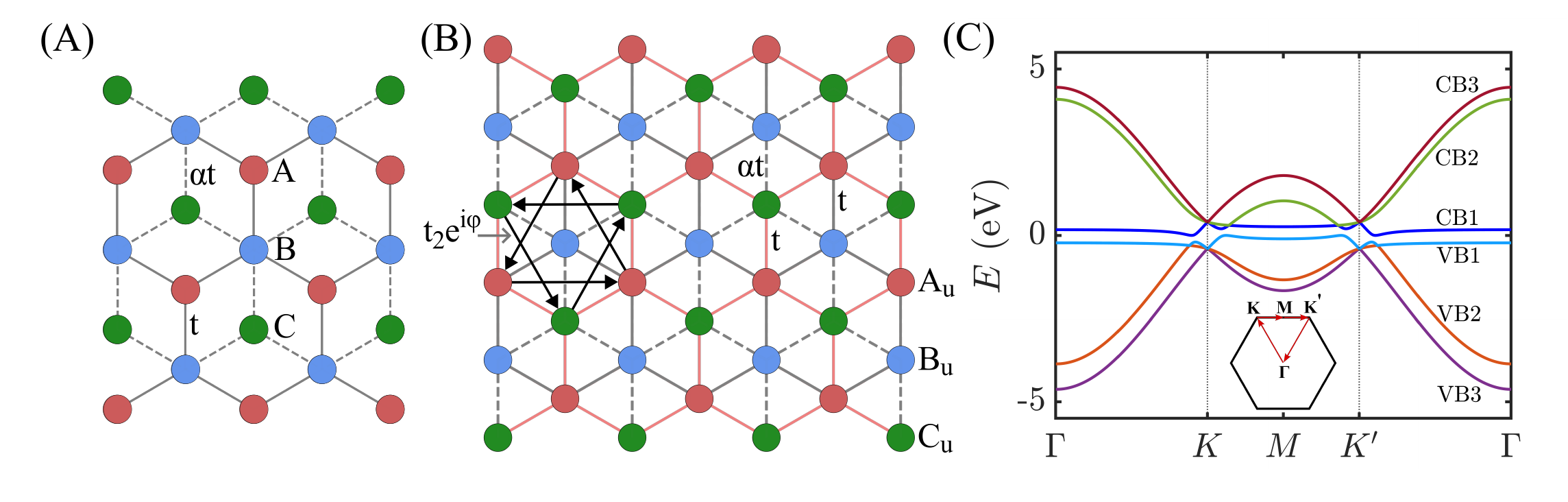}
    \caption{(A) Schematic diagram of single layer $\alpha-\mathcal{T}_3$ lattice. (B) Schematic diagram (top view) of cyclically stacked bilayer $\alpha-\mathcal{T}_3$ lattice, positions of sites $A_u(C_l)$, $B_u(A_l)$ and $C_u(B_l)$ of the upper (lower) layer denoted by red, green and blue respectively. The NN hopping of the upper and lower layers are denoted as grey and red lines, and the hopping strength between sites $B$ and $C$ are shown as $\alpha t$ (dashed) and between $B$ and $A$ (solid) as $t$. The next NNN complex hopping of strength $t_2 e^{i\phi}$ for $A$ and $C$ sublattices are shown with black arrows for anti-clockwise direction. (C) Band structure of cyclically $(A_lB_u - B_lC_u - C_lA_u)$ stacked bilayer $\alpha-\mathcal{T}_3$ lattice, calculated along the path of high symmetry points $(\Gamma - K - M - K^\prime - \Gamma)$ in the absence of Haldane flux and the path in the hexagonal Brillouin zone (inset).}
    \label{fig: schematic}
\end{figure*}
In addition to TBG, flat bands are observed in Kagome \cite{PhysRevB.62.R6065, PhysRevB.80.113102}, $\mathcal{T}_3$ or dice \cite{PhysRevB.34.5208, PhysRevLett.81.5888}, and Lieb lattices \cite{PhysRevB.82.085310, PhysRevA.83.063601}. The $\mathcal{T}_3$ or dice lattice exhibits flat bands near charge neutrality due to destructive interference of wave functions, resulting in electronic bands without dispersion \cite{Liu_2014, PhysRevB.99.045107, PhysRevB.78.125104, PhysRevMaterials.5.084203}. These flat bands are trivial in nature with non-singular Bloch wave functions \cite{Liu_2014, PhysRevB.99.045107, PhysRevB.78.125104, PhysRevMaterials.5.084203}. Experimental realization of the dice lattice was proposed and discussed using an optical lattice, utilizing three pairs of counter-propagating identical laser beams with the same wavelength of $\lambda = \frac{3}{2a}$, where $a$ is the lattice constant, and a Josephson Junction array \cite{PhysRevB.73.144511, PhysRevA.80.063603}. Followed by another proposal for the experimental realization of the dice lattice using artificial heterostructures of trilayer cubic lattices grown along the (111) direction, such as $SrTiO_3/SrIrO_3/SrTiO_3$ \cite{PhysRevB.84.241103}. $\alpha$-$\mathcal{T}_3$ shows a continuously variable Berry phase from $\pi$ (graphene) to $0$ (dice) with the variable $\alpha$, and the Berry phase dependence of quantized Hall conductivity, dynamical longitudinal optical conductivity, and SDH oscillation has been established \cite{PhysRevB.92.245410}. The effect of the variable Berry phase on the orbital magnetic susceptibility of $\alpha$-$\mathcal{T}_3$ with the variation of $\alpha$ is observed \cite{PhysRevLett.112.026402}. Further, the magnetotransport property of $\alpha$-$\mathcal{T}_3$ has been studied, and it is found that the Hall conductivity undergoes a smooth transition from $\sigma_{yx} = 2(2n+1)(e^2/h)$ to $\sigma_{yx} = 4n(e^2/h)$ with $n = 0,1,2$... as $\alpha$ is tuned from $0$ to $1$ \cite{biswas2016magnetotransport}. The effect of disorder and staggered lattice potential on integer quantum Hall plateaus has been established \cite{PhysRevB.102.235414}. Floquet states and the variable Berry phase-dependent photoinduced gap in the $\alpha$-$\mathcal{T}_3$ lattice irradiated by circularly polarized on-resonant light have been studied \cite{PhysRevB.98.075422}. Later, the floquet topological phase transition in a single-layer $\alpha$-$\mathcal{T}_3$ lattice is shown by breaking of time-reversal symmetry (TRS) using off-resonant circularly polarized light and varying $\alpha$ discovered the topological phase transition ($C_n = 2$ to $C_n = 1$) at $\alpha = \frac{1}{\sqrt{2}}$ \cite{PhysRevB.99.205429}. More recently, higher Chern numbers have been observed in single-layer Dice-lattice. A Haldane-like model on the Dice lattice (single-layer) showed Chern numbers of the three bands $\pm2$ (dispersive VB and CB) and $0$ (flat band in the middle). Also, the quantum anomalous hall effect (QAHE) with two chiral channels per edge was observed \cite{PhysRevB.101.235406}. A topological phase transition from Chern insulating ($\sigma_{xy} = 2e^2/h$) to a trivial ($\sigma_{xy} = 0$) insulating phase is observed at the semi-Dirac limit of the Dice lattice \cite{PhysRevB.107.035421}.

However, the flat bands near the charge neutrality in $\alpha$-$\mathcal{T}_3$ lattice become dispersive at the band touching high-symmetric point ($K$ and $K^{\prime}$) in the presence of spin-orbit coupling and become non-trivial \cite{PhysRevB.84.241103}. Interestingly, when two layers of $\alpha$-$\mathcal{T}_3$ lattice stack on top of each other to form a bilayer with four non-equivalent commensurate stacking, the effective low-energy model shows the dispersive nature of the flat bands near the charge neutrality \cite{PhysRevB.108.075166}. The topological properties of these nearly flat bands of bilayer $\alpha$-$\mathcal{T}_3$ lattices have not been explored. 

This paper is organized in the following way. In section (II) we presented the model Hamiltonian of the bilayer system. In Sec. (III) we discuss the result in which Sec. (III)A shows the spectral properties for different stacking configurations. Sec. (III)B and (III)C present the topological properties and show the Chern phase diagrams. Further in Sec. (III)D, we present the anomalous hall conductivity. We finally conclude with a brief summary of the results obtained in Sec. (IV).

\begin{figure*}[!ht]
    \centering
    \includegraphics[scale=0.4]{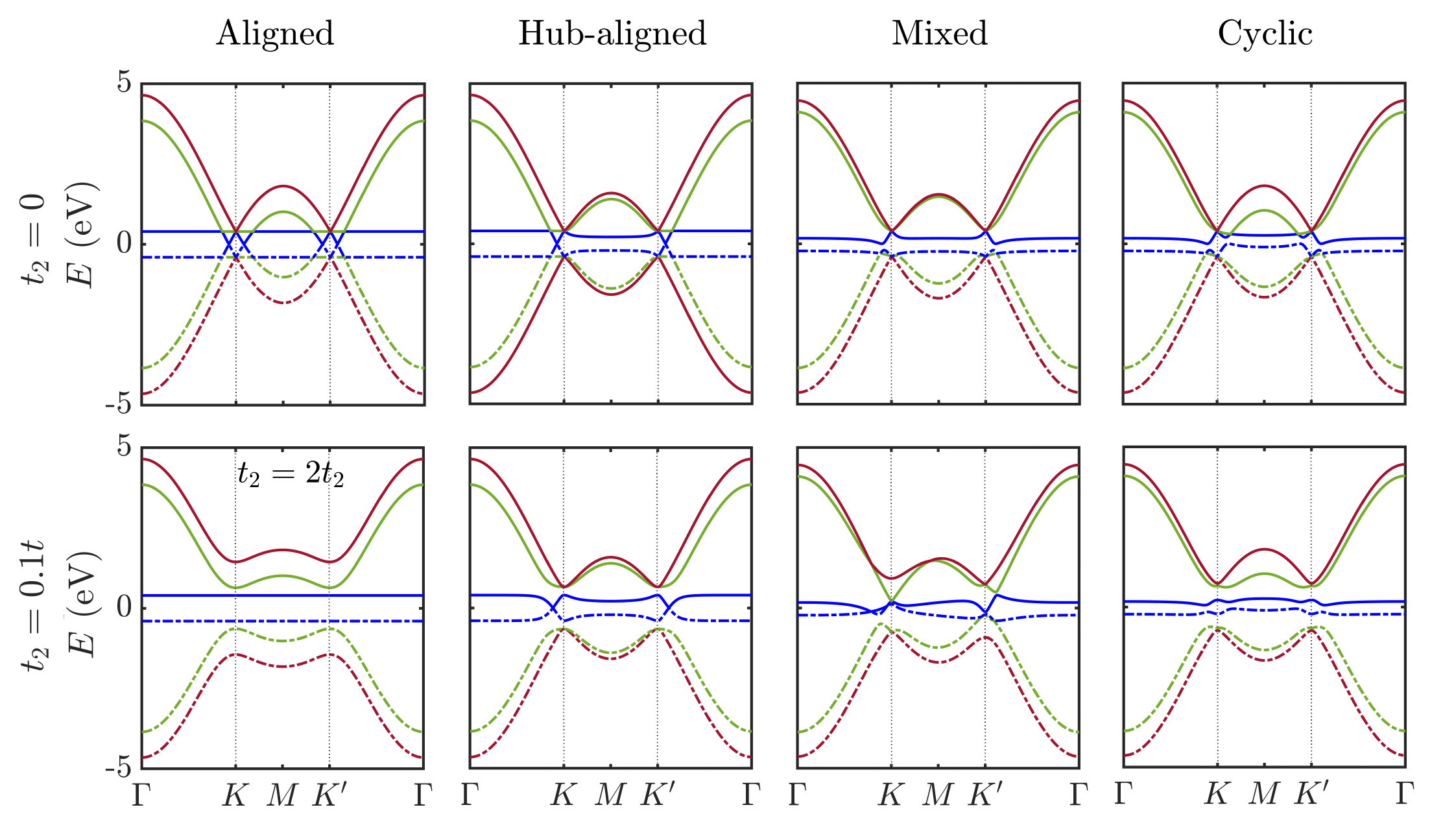}
    \caption{Electronic band structure of all four nonequivalent commensurate stackings of bilayer $\alpha-\mathcal{T}_3$ lattice, i.e Aligned $(A_lA_u - B_lB_u - C_lC_u)$, Hub-aligned $(A_lC_u - B_lB_u - C_lC_u)$, Mixed $(A_lA_u - B_lC_u -C_lB_u)$ and cyclic $(A_lB_u - B_lC_u - C_lA_u)$ for zero and non-zero values of $t_2$.}
    \label{fig: band_structures}
\end{figure*}

\section{The model Hamiltonian}
\subsection{single layer and bilayer \texorpdfstring{$\alpha-\mathcal{T}_3$} lattice}
The $\alpha-\mathcal{T}_3$ lattice is similar to a honeycomb lattice but with an additional lattice site at the center of the hexagon, as shown in Figure \ref{fig: schematic}(A). The two sub-lattice sites, $A$ and $B$, form the honeycomb lattice with a nearest-neighbor (NN) hopping amplitude of $t$ (solid line). Additionally, the third sub-lattice site, $C$, located at the center of the hexagon, connects to site $B$ with NN hopping, and the hopping amplitude can be tuned with a parameter $\alpha$, defined as $\alpha t$ (dashed line). This lattice structure is bipartite, with sites $A$ and $C$ referred to as 'rim-atom' sites, and the $B$ lattice sites are known as 'hub-atom' sites. The rim-atoms have 3 nearest neighbors, whereas hub-atoms have 6 nearest neighbors \cite{PhysRevB.34.5208,PhysRevLett.81.5888,PhysRevB.63.134503,PhysRevB.73.144511,PhysRevA.80.063603,PhysRevB.84.115136,PhysRevB.93.165433,PhysRevB.88.161413}.
The real parameter $\alpha$ smoothly interpolates between two limiting cases of the model, $\alpha = 0$ and $1$, referred to as graphene and the dice lattice, respectively. The intermediate values, $0 < \alpha < 1$, are defined as the $\alpha-\mathcal{T}_3$ lattice \cite{PhysRevLett.112.026402}. In this paper, we discuss the bilayer structure of the $\alpha-\mathcal{T}_3$ lattices. The $\alpha-\mathcal{T}_3$ bilayer has four non-equivalent vertically aligned commensurate stackings \cite{PhysRevB.108.075166}. The two layers stack on top of each other, where each sub-lattice ($A_u, B_u, C_u$) from the top layer aligns exactly on top of each sub-lattice ($A_l, B_l, C_l$) of the bottom layer, is called Aligned ($A_lA_u - B_lB_u - C_lC_u$). A $\pi/3$ rotation of the vertical axis passing through the aligned hub-atoms ($B_lB_u$) in the 'Aligned' bilayer gives rise to another stacking with sub-lattice alignment changed to ($A_lC_u - B_lB_u - C_lA_u$), where the two hub-atoms ($B_l$ and $B_u$) remain aligned, and rim-atoms' alignment is exchanged, called 'Hub-aligned' stacking. Similarly, a $\pi/3$ rotation along any of the aligned rim-atoms ($A_lA_u$ or $C_lC_u$) of the 'Aligned' bilayer gives rise to different stackings where one of the aligned rim-atoms remains unchanged and the hub-atom and the second rim-atoms have mixed alignment ($A_lA_u - B_lC_u - C_lB_u$) or ($A_lC_u - B_lA_u - B_lB_u$), called 'Mixed' stacking. The two configurations of 'Mixed' stacking are equivalent structures as both rim atoms are equivalent. Additionally, sliding one of the layers in the 'Aligned' stacking for a distance $a_{0}$, which is the length between two sublattices, gives rise to the cyclic alignment of the sub-lattices ($A_lB_u - B_lC_u - C_lA_u$), called 'Cyclic' stacking. The discrete symmetries of all four bilayer stacking arrangements are discussed thoroughly in Ref.\cite{PhysRevB.108.075166}. The single-layer of the dice lattice preserves the in-plane inversion symmetry with the interchange of two rim sites, which can be shown with the $3\times3$ matrix, $W_0$, 
\begin{equation}
    W_0 = \begin{pmatrix}
        0 & 0 & 1\\
        0 & 1 & 0\\
        1 & 0 & 0
    \end{pmatrix}
    \label{eq:inv_symmetry_matrices}
\end{equation}

In the case of the $\alpha-\mathcal{T}_3$ lattice bilayer within the dice lattice limits, the Aligned and Hub-aligned stacking preserves in-plane inversion ($W_1$) and full inversion ($W_2$) symmetries with symmetry matrices $W_1 = \mathds{1} \otimes W_0$ and $W_2 = \tau_x \otimes W_0$, where $\tau_x$ is the vector of Pauli spin matrix defined in the layer space. $W_1$ and $W_2$ are $6 \times 6$ matrices which correspond to an interchange of rim sites in both planes and an additional interchange of layers, respectively. However, the Cyclic stacking preserves only $W_2$ symmetry \cite{PhysRevB.108.075166}. When $\alpha \ne 1$, for general $\alpha-\mathcal{T}_3$ cases, both $W_1$ and $W_2$ inversion symmetries are no longer preserved. The coupling between hub and rim atoms of different layers leads to particle-hole symmetry breaking in the case of mixed and cyclic stacking, as shown in later sections.
\begin{figure*}[!ht]
    \centering
    \includegraphics[scale=0.4]{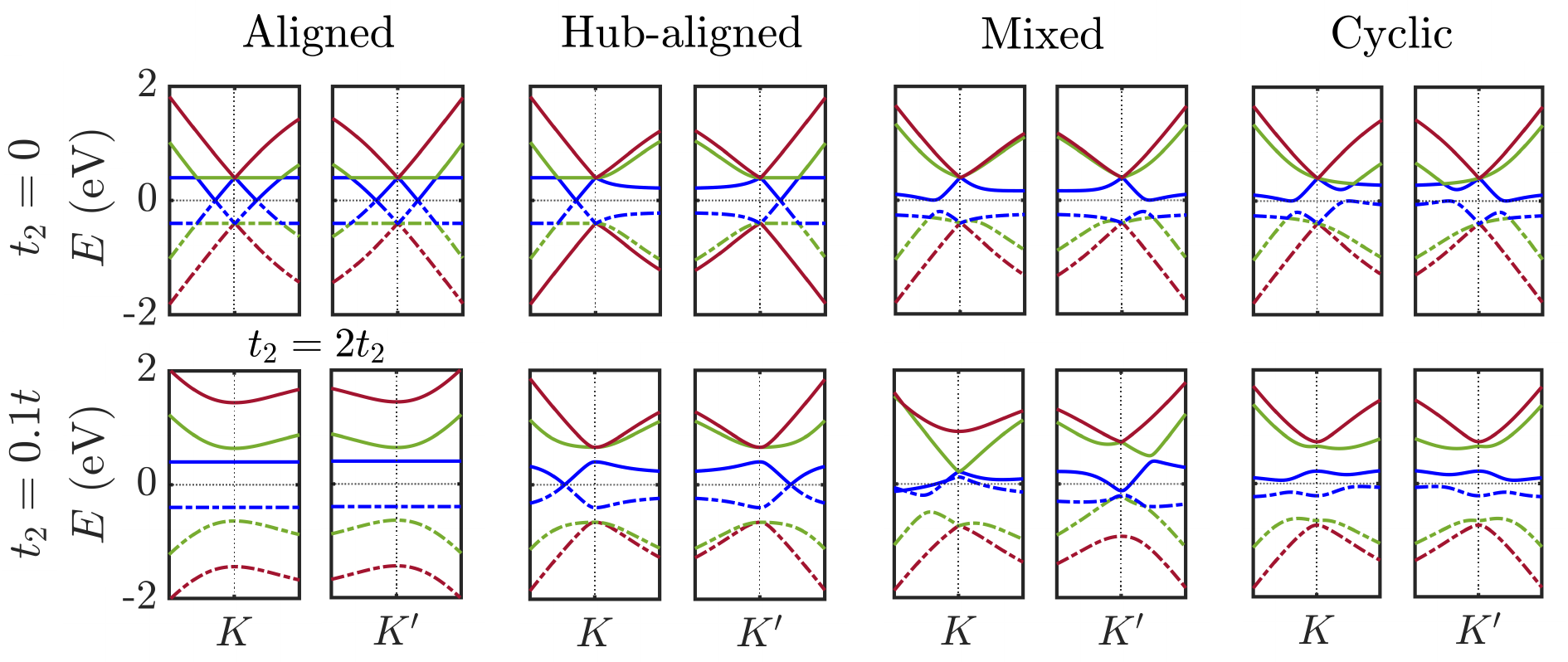}
    \caption{Zoomed figure of the spectrum at the Dirac points $K$ and $K^\prime$. The dotted lines show the Fermi energy.}
    \label{fig:band_structure_zoomed}
\end{figure*}

\subsection{Tight-binding Hamiltonian}
The tight-binding Hamiltonian for bilayer $\alpha-\mathcal{T}_3$ lattice can be written as
\begin{multline}
    H = \sum_{m\in l,u}\biggl[ \sum_{\langle i,j \rangle} t_{ij} C_i^{m\dagger}C_j^m + t_2 \sum_{\langle \langle i,j \rangle \rangle} e^{i\phi_{ij}^m}C_i^{m^\dagger}C_j^m +\\
     \sum_i \Delta_i C_i^{m\dagger}C_i^m + h.c \biggr] + \biggl[ t_{\perp}\sum_{\langle p,q \rangle} C_p^{l\dagger}C_q^u + h.c\biggr]
     \label{eq:Hamiltonian_expression}
\end{multline}

Where $m$ is the layer index, which takes $l$ ($u$) for the upper (lower) indices, and $C_i^{m\dagger}$ ($C_i^m$) is the creation (annihilation) operator of the $i^{th}$ site of layer $m$.  We used $t = -1$ eV, $t_2 = 0.1t$, and $t_\perp = -0.4$ eV. The first term implies the nearest neighbor hopping along directions $\boldsymbol{\delta_1} = (0, a_{0})$, $\boldsymbol{\delta_2} = \bigl(-\frac{\sqrt3 a_{0}}{2}, -\frac{a_{0}}{2}\bigr)$, and $\boldsymbol{\delta_3} = \bigl(\frac{\sqrt3 a_{0}}{2}, -\frac{a_{0}}{2}\bigr)$ with strength $t$ among $A$ and $B$ sublattices and $\alpha t$ among $B$ and $C$ sublattices. The second term is the complex next nearest neighbor (NNN) hopping (Haldane term) along the direction of the three vectors $\boldsymbol{\nu_1} = \boldsymbol{\delta_2} - \boldsymbol{\delta_3}$, $\boldsymbol{\nu_2} = \boldsymbol{\delta_3} - \boldsymbol{\delta_1}$, and $\boldsymbol{\nu_3} = \boldsymbol{\delta_1} - \boldsymbol{\delta_2}$, with hopping strength $t_2$ and phase $\phi_{ij}^m$, which takes positive values for the electrons hopping clockwise and negative for anticlockwise hopping for each layer $m$. In the present work, we consider the Haldane Phase to be same for both the layers as $\phi_{ij}^m = \phi_{ij}$. The NNN hopping from the $B$ sublattice (hub-atom) along that direction is forbidden due to the high potential barrier. The third term is the onsite energy term that takes positive and negative values for $A_m$ and $C_m$ sublattices, respectively. The last term in the Hamiltonian is the interlayer coupling term with coupling strength $t_{\perp}$. We considered $t_{\perp}$ to have the same values for coupling among different sublattices. The Fourier transform of the Hamiltonian (Eq. \eqref{eq:Hamiltonian_expression}) takes the form of,

\begin{equation}
    H =  \begin{pmatrix}
        H^l & H_c\\
        H_c^T & H^u
    \end{pmatrix}
    \label{eq: bilayer_hamiltonian_matrix}
\end{equation}

Here, $H^l$ and $H^u$ are $3\times 3$ Hamiltonian matrices for the lower and upper layers, respectively, with the sublattice basis $A_l$, $B_l$, $C_l$ for the bottom layer, and $A_u$, $B_u$, $C_u$ for the top layer

\begin{equation}
H^l = H^u = \begin{pmatrix}
        f_z^+(k) & f(k,t) & 0\\
        f^*(k,t) & 0 & f(k,\alpha t)\\
        0 & f^*(k,\alpha t) & f_z^-(k)
    \end{pmatrix}
    \label{eq:single_hamiltonian}
\end{equation}

and the general form of the coupling Hamiltonian
\begin{equation}
H_c = \begin{pmatrix}
    t^{am}_{\perp} & t^{c}_{\perp} & t^{h}_{\perp}\\
    0 & t^{ah}_{\perp} & t^{mc}_{\perp}\\
    t^{hc}_{\perp} & t^{m}_{\perp} & t^{a}_{\perp}
\end{pmatrix}
\end{equation}

Where, $f_z^{\pm} = f_1(k) \pm f_2(k)$ and $f(k,t^\prime) = f_x(k,t^\prime) - if_y(k,t^\prime)$, and $t^\prime$ takes the values of either $t$ or $\alpha t$.
\begin{multline*}
     f_x(k,t^\prime) = t^\prime\bigg\{\cos a_{0}k_y+2\cos\frac{\sqrt{3}a_{0}k_x}{2}
     \cos\frac{a_{0}k_y}{2}\bigg\},
\end{multline*}
\begin{multline*}
    f_y(k,t^\prime) = t^\prime\bigg\{\sin a_{0}k_y - 2\cos\frac{\sqrt3a_{0}k_x}{2}
     \sin \frac{a_{0}k_y}{2}\bigg\},
\end{multline*}

\begin{multline*}
    f_1(k) = 2t_2\cos\phi\bigg\{ 2\cos\frac{\sqrt3 a_{0}k_x}{2} \cos\frac{3a_{0}k_y}{2}\\
    +\cos\sqrt3 a_{0}k_x \bigg\}
\end{multline*}

\begin{multline*}
    f_2(k) = \Delta - 2t_2\sin{\phi}\bigg\{ 2\sin\frac{\sqrt{3}a_{0}k_x}{2}\cos\frac{3a_{0}k_y}{2}  \\
    - \sin\sqrt3 a_{0}k_x \bigg\},
\end{multline*}
All the $t_{\perp}$ terms vanishes except $t^{am}_{\perp}$, $t^{ah}_{\perp}$ and $t^a_{\perp}$ for "Aligned" case, $t^{hc}_{\perp}$, $t^{ah}_{\perp}$ and $t^h_{\perp}$ for "Hub-aligned", $t^{am}_{\perp}$, $t^m_{\perp}$ and $t^{mc}_{\perp}$ for "Mixed", and $t^{hc}_{\perp}$, $t^c_{\perp}$ and $t^{mc}_{\perp}$ for "Cyclic" stacking cases respectively.
Hamiltonian $H_l(H_u)$ for lower (upper) layer in terms of spin-1 matrices is given as: 
 \begin{equation}
     H_t = H_b = f_1(k)S_0 + f_2(k)S_z + f_x(k,t^\prime)S_x + f_y(k,t^\prime)S_y   
 \end{equation}
 where,
 \begin{equation}
   S_x = \begin{pmatrix}
     0 & 1 & 0\\
     1 & 0 & 1\\
     0 & 1 & 0
 \end{pmatrix},  \hspace{0.1cm}
 S_y=i\begin{pmatrix}
     0 & -1 & 0\\
     1 & 0 & -1\\
     0 & 1 & 0
 \end{pmatrix} \nonumber
 \end{equation}
 \begin{equation}
 S_z = \begin{pmatrix}
     1 & 0 & 0\\
     0 & 0 & 0\\
     0 & 0 & -1
 \end{pmatrix},\hspace{0.3cm} 
 S_0 = \begin{pmatrix}
     1 & 0 & 0\\
     0 & 0 & 0\\
     0 & 0 & 1
 \end{pmatrix} \nonumber
 \end{equation}
 In the absence of next-nearest-neighbor (NNN) hopping, this Hamiltonian can be linearized for momentum around the Dirac points, taking the form of a pseudospin-1 Dirac-Weyl Hamiltonian \cite{PhysRevA.80.063603},
 \begin{equation}
     H_\xi (q) = \hbar v_f (\xi q_xS_x + q_y S_y) 
 \end{equation}
 where $\textbf{q} = \textbf{k} - \textbf{K}$, momentum around Dirac points, $\xi = \pm 1$ are the valley indices, $v_f$ is Fermi velocity.
 
\begin{figure}[!hb]
    \centering
    \includegraphics[width=0.4\textwidth]{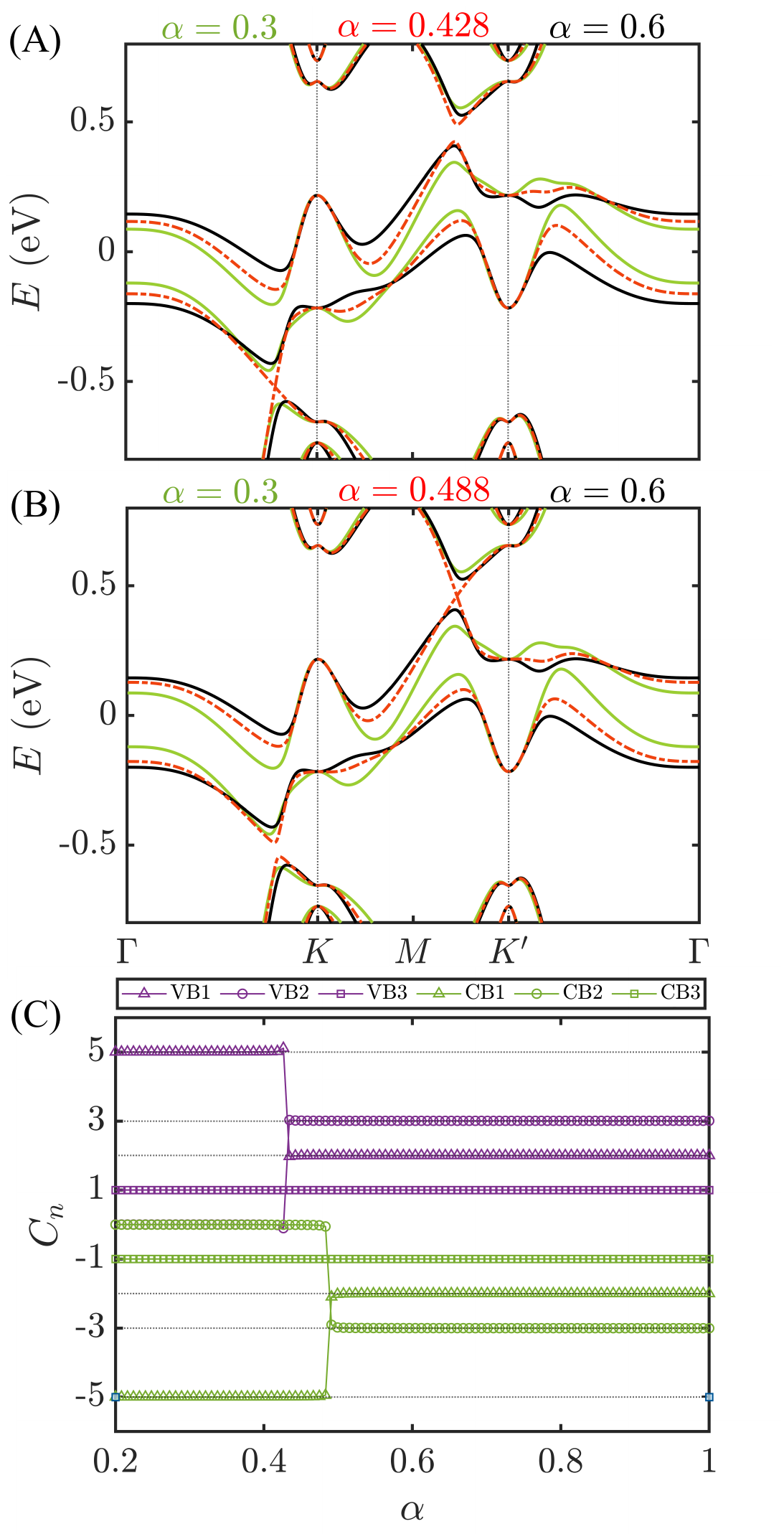}
    \caption{Band structure of cyclic bilayer $\alpha-\mathcal{T}_3$ for different values of $\alpha$ showing band closing and opening for (A)  VB, $\alpha = 0.3$ (green), $\alpha = 0.428$ (red) and $\alpha = 0.6$ (black) (B) CB $\alpha = 0.3$ (green), $\alpha = 0.428$ (red) and $\alpha = 0.6$ (black), for all three cases we use $t_2 = 0.1t$, $\phi =\frac{\pi}{2}$, $\Delta = 0$. The bands are no longer symmetric under the exchange of valley for $\alpha \ne 1$. (C) The variation of Chern number as a function of a $\alpha$ for three VB (purple) and CB (green).}
    \label{fig:phase_transition_line_plt}
\end{figure}

\section{Result and Discussions}
\subsection{Electronic spectrum}
Now, we discuss the energy spectrum of single-layer and bilayer $\alpha-\mathcal{T}_3$ lattices of different stackings, obtained using the tight-binding Hamiltonian discussed above under the influence of Haldane flux. The band structures are calculated along the high-symmetry points: $\Gamma(0,0)$, $K\left(-\frac{4\pi}{3\sqrt{3}a_0},\frac{2\pi}{3a_0}\right)$, $M\left(0,\frac{2\pi}{3a_0}\right)$, and $K^\prime\left(\frac{4\pi}{3\sqrt{3}a_0},\frac{2\pi}{3a_0}\right)$ in the hexagonal Brillouin zone (BZ). In the single-layer $\alpha-\mathcal{T}_3$ lattice, the three sublattices per unit cell result in three bands in the energy spectrum: one non-dispersive flat band at zero energy and two conical bands from the valence band (VB) and conduction band (CB). All three bands intersect each other at $K$ and $K^{\prime}$ with the flat band near the charge neutrality, similar to graphene. We introduced the Haldane terms in the Hamiltonian and the three-fold degeneracy is lifted, opening a gap at the Dirac points ($K$ and $K^{\prime}$), as shown in Figure \ref{fig:single_layer}(a). Furthermore, we varied $\alpha$ from 1 to 0.2, which makes the flat bands more dispersive. It is worth noting that the flat band energy remains zero at the two Dirac points and shows dispersion in the remaining high-symmetry path in the BZ. 
The bilayer $\alpha-\mathcal{T}_3$ lattice's energy spectrum appears to be sensitive to the choice of the coupling strengths among the different sublattices of the two layers. All four non-equivalent commensurate stackings for the dice lattice limit ($\alpha = 1$) with and without the inclusion of the Haldane flux are shown in.\ref{fig: band_structures}. In the bilayer geometry, the inclusion of the Haldane flux breaks time-reversal symmetry (TRS) in all four stackings of the bilayer $\alpha-\mathcal{T}_3$ lattice. Interestingly, particle-hole symmetry is preserved in Aligned and Hub-aligned stacking due to the hub-hub and rim-rim interlayer interactions, while it is broken in the Mixed and Cyclic stacking due to the presence of hub-rim interactions between the two layers. Furthermore, the low-energy bands dynamically change with increasing strengths of the next-nearest-neighbor (NNN) complex hopping strength ($t_2$), as shown in Figure \ref{fig:band_structure_zoomed}. These changes are symmetric at both the Dirac points ($K$ and $K^{\prime}$).
In the case of Aligned stacking, the valence and conduction bands degenerate at the Dirac point and around the Dirac points, while the flat bands are pushed away from zero energy.
For a large value of NNN complex hopping strength ($t_2$), the degeneracy among all six bands is lifted, and the flat bands are around zero energy. Similar to the single layer\cite{PhysRevB.107.035421}, the presence of NNN complex hopping strength ($t_2$) does not make them non-trivial, and their Chern numbers remain zero. In the Hub-aligned and Mixed stacking cases, the bands are strongly degenerate and are not separated with varying values of the NNN complex hopping strength ($t_2$). The most interesting case is the cyclic stacking of the bilayer $\alpha-\mathcal{T}_3$ lattice. In this stacking, the conduction and valence bands are separated at charge neutrality with $t_2 = 0$, and all the bands become separated with the inclusion of Haldane flux with $t_2 = 0.1t$. The Haldane flux breaks the time-reversal symmetry of the system, and all six bands become separated, leaving a finite gap at zero energy, which makes the system insulating. The energy spectrum of cyclic stacking consists of two isolated, nearly-flat bands near the Fermi level and four dispersive higher-energy bands, two in each conduction and valence band. The bandwidth of the two nearly-flat bands (0.19 eV for valence and 0.21 eV for conduction) is found to decrease (0.16 eV for valence and 0.19 eV for conduction) for $t_2 = 0.1$ eV and becomes more flattened for $t_2 = 0.2$ eV. These flat bands are intriguing in a simple bilayer geometry when compared to complex systems like twisted bilayer graphene. Furthermore, we found that these flat bands are non-trivial, with non-zero Chern numbers, making the cyclic-stacked bilayer $\alpha-\mathcal{T}_3$ lattice a potential candidate to study topological properties. Additionally, by changing the hopping strength $\alpha t$ (see Figure \ref{fig: schematic}(B), the inversion symmetry is broken in the system, causing the two nearly flat bands to become dispersive.

\begin{figure*}[!ht]
    \centering
    \includegraphics[scale=0.4]{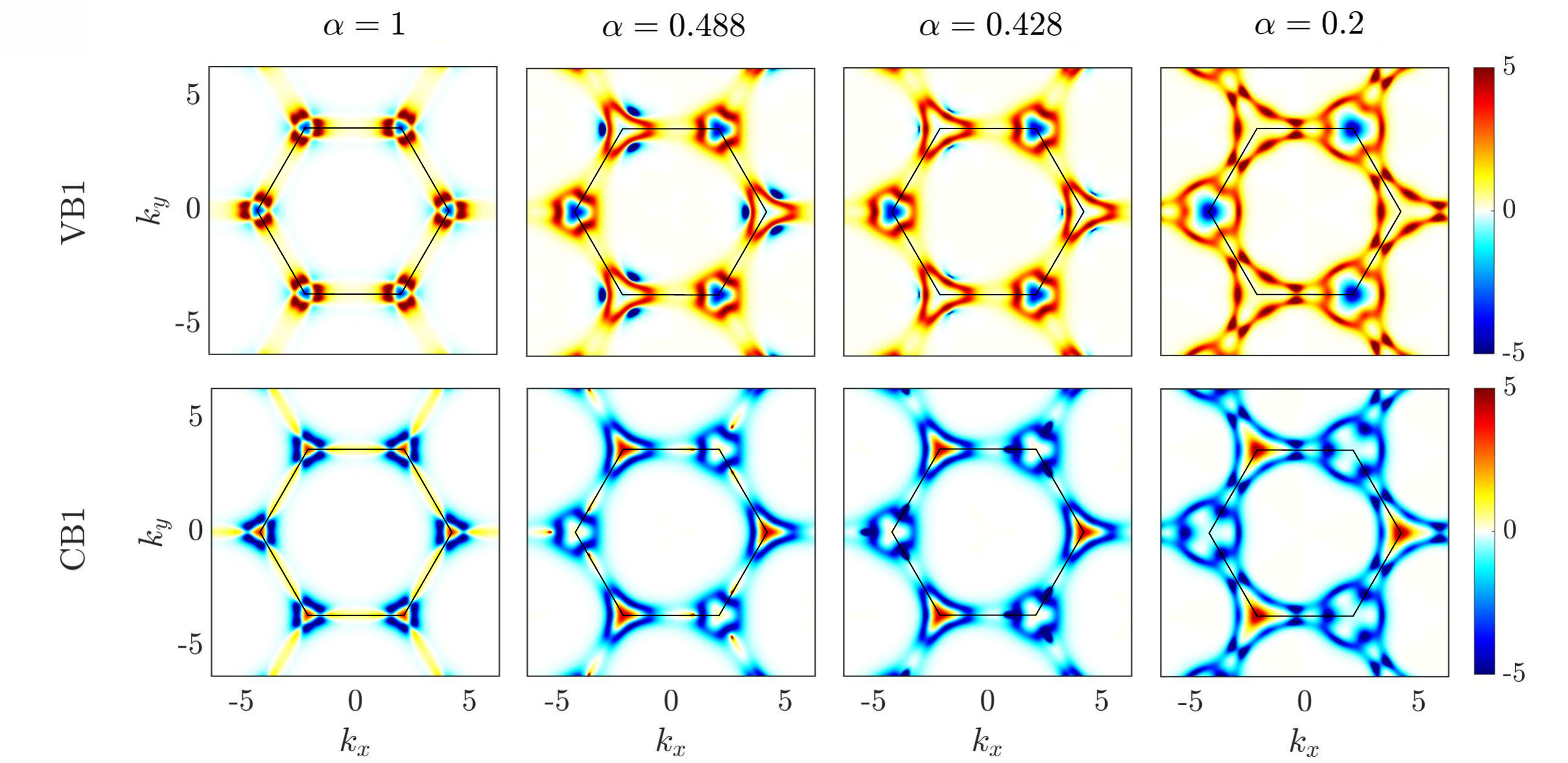}
    \caption{Berry curvature distribution in the $k_x - k_y$ plane for the nearly flat valance (VB1) and conduction (CB1) bands for $\alpha = 1$ (dice lattice limit), $\alpha = 0.488$ (at CB phase transition), $\alpha = 0.428$ (at VB phase transition) and $\alpha = 0.2$. Here $t_2$, $\Delta$ and $\phi$ are fixed at $0.1t$, $0$ and $\frac{\pi}{2}$ respectively.}
    \label{fig:Berry_curvature}
\end{figure*}

\subsection{Topological properties}
Furthermore, we have conducted an extensive study of the cyclic stacking to investigate the topological properties of the bilayer $\alpha-\mathcal{T}_3$ lattice. We calculated the Berry curvatures for the energy spectrum in the Brillouin Zone (BZ) for all six isolated bands and determined their respective Chern numbers. The z-component of the Berry curvature was calculated numerically using\cite{RevModPhys.82.1959}
\begin{equation}
    \Omega_n(k_x,k_y) = -2\sum_{n^\prime \ne n} Im [\frac {\bra{u_n} \frac{\partial H}{\partial k_x} \ket{u_{n^\prime}} \bra{u_{n^\prime}} \frac{\partial H}{\partial k_y} \ket{u_n}} {(E_{n^\prime} - E_n)^2} ]
\end{equation}
 for the $n^{th}$ band for each $k$-point, summing over all neighboring $n^\prime$ bands. Here, $\ket{u_n}$ are the block states, and $E_n$ are the eigenvalues of the Hamiltonian given in Eq. \eqref{eq: bilayer_hamiltonian_matrix} for the $n$-th band. The surface integral of the Berry curvature over the first BZ yields $2\pi C_n$ for the $n$-th band, where $C_n$ is an integer called the Chern number or TKNN index \cite{PhysRevLett.49.405, PhysRevLett.51.51}. The complex NNN hopping term plays a crucial role in determining the topological properties of the cyclic-stacked $\alpha-\mathcal{T}_3$ bilayer lattice. In the bilayer cyclic $\alpha-\mathcal{T}_3$ lattice, all six bands become topologically non-trivial. These include the flat bands VB1 and CB1, and the higher-energy dispersive bands VB2, VB3, CB2, and CB3 (see Fig. 1(C)). We began from the dice lattice limit and varied $\alpha$ from 1 to 0.2 to observe the topological evolution in the system. At the dice lattice limit, the system becomes topological due to non-zero Chern numbers. VB1 (CB1), VB2 (CB2), and VB3 (CB3) possess Chern numbers of +2 (-2), +3 (-3), and +1 (-1), respectively. The continuous variation of $\alpha$ shows a phase transition from one topological insulating phase to another through a band crossing between VB1 and VB2 in the valence band at $\alpha = 0.428$ between $\Gamma$ and $K$ points and between CB1 and CB2 in the conduction band at $\alpha = 0.488$ between $M$ and $K^\prime$ points. After the transition, the nearly flat bands (VB1 and CB1) acquire very high Chern numbers of ±5, while the middle bands (VB2 and CB2) become topologically trivial with a Chern number of 0. The Chern numbers for the highest energy bands (VB3 and CB3) remain unchanged, as expected, as these bands do not take part in the band crossing. Fig. \ref{fig:phase_transition_line_plt}(A) and Fig. \ref{fig:phase_transition_line_plt}(B) show the dispersive nearly flat band structures for three different values of $\alpha$ aside from 1. The red dashed curve shows the band crossing at (A) the valence band and (B) the conduction band. Fig. \ref{fig:phase_transition_line_plt}(C) shows the Chern number line plots as a function of $\alpha$ for all six bands, clearly showing the switching of Chern numbers at the transition points. This behavior is not observed for its single-layer counterpart in the presence of the Haldane term. However, a similar study of topological phase transition has been observed for the single-layer $\alpha-\mathcal{T}_3$ lattice when irradiated with off-resonant circularly polarized light. The variation in $\alpha$ breaks the $W_2$ inversion symmetry, as visualized in the Berry curvature distribution shown in Fig. \ref{fig:Berry_curvature}. The Berry curvature distribution over the $k_x-k_y$ plane at the dice lattice limit ($\alpha = 1$) exhibits the same features at $K$ and $K^\prime$ points, reflecting the preserved $W_2$ inversion symmetry in the system. For $\alpha$ values other than 1, the symmetry breaks, resulting in different Berry curvature distributions at $K$ and $K^\prime$. 

\begin{figure}[!hb]
    \centering
    \includegraphics[width=0.4\textwidth]{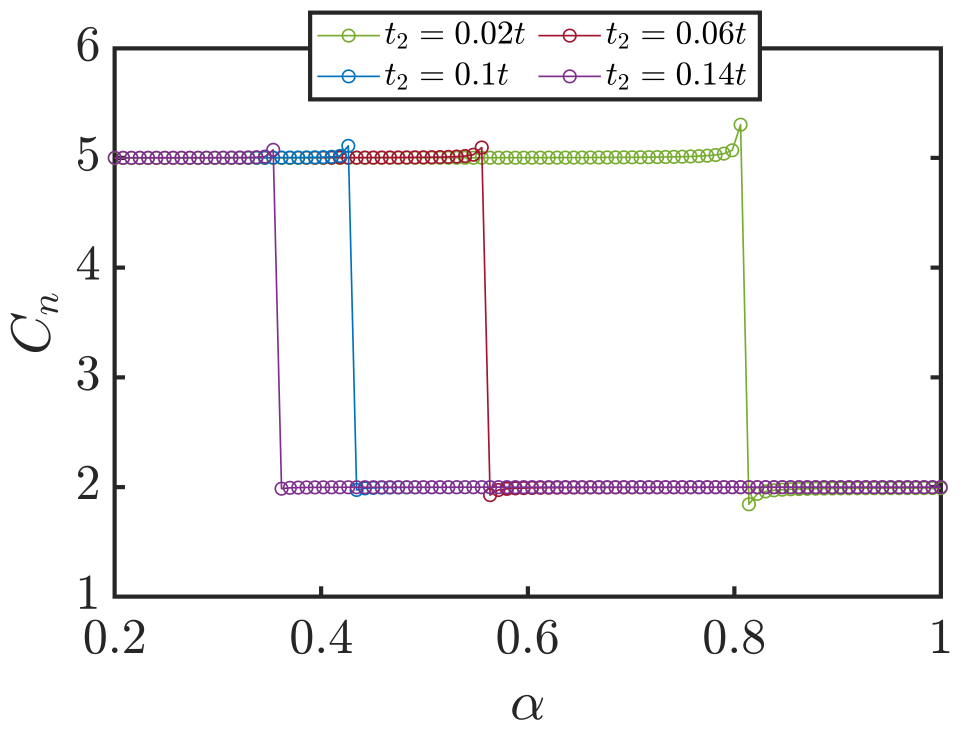}
    \caption{Chern number variation of VB1 $(n=3)$ with $\alpha$ for different values of NNN hopping parameter $t_2$. We kept $\Delta = 0$ and $\phi = \frac{\pi}{2}$ here.}
    \label{fig:ch_no_line_plt_t2}
\end{figure}

\subsection{Chern phase diagrams}
\begin{figure}
    \centering
    \includegraphics[width=0.48\textwidth]{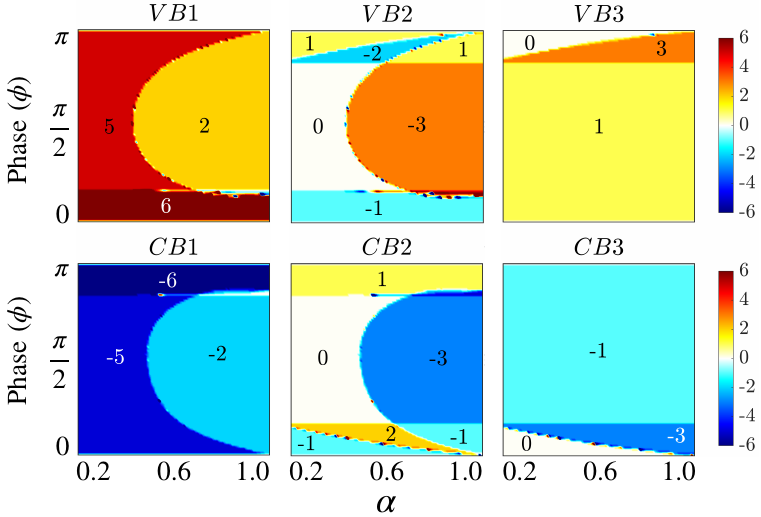}
    \caption{Chern number phase diagram of three valance bands (left column) and conduction bands (right column) of Cyclic bilayer $\alpha-\mathcal{T}_3$ lattice in the parameter space $\alpha$ $(0.2 \sim 1)$ and $\phi$ $(0 \sim \pi)$. $t_2 = 0.1t$, $\Delta = 0$ for all bands.}
    \label{fig:alpha_phi_phase}
\end{figure}

The bilayer cyclic $\alpha-\mathcal{T}_3$ lattice exhibits a rich interplay between time-reversal and inversion asymmetry controlled by the flux phase $\phi$ and the parameter $\alpha$, respectively. The simultaneous variation of $\phi$ and $\alpha$ gives rise to a multitude of new topological phases. In Fig. \ref{fig:phase_transition_line_plt}, we demonstrate how the Chern insulating phase evolves as we fix $\phi = \frac{\pi}{2}$ while varying $\alpha$. However, there is more to explore within the parameter space $0 < \phi < \pi$. Our investigations reveal additional Chern phases, which we have summarized in the Chern number phase diagram shown in Fig.\ref{fig:alpha_phi_phase}. Here, we examine the dependencies of the Chern numbers for $VB1 (CB1)$, $VB2 (CB2)$, and $VB3 (CB3)$ on $\alpha$ and $\phi$. Here we change the Haldane Phase for bothe layers simultaneously. Notably, the Chern numbers of the lower energy valance/conduction bands exhibit sensitivity to both $\alpha$ and $\phi$. Meanwhile, the second set of dispersive bands (VB3 and CB3) display a negligible dependence on $\phi$ within a very small range. In the phase diagram, we also present the conventional Haldane phase diagram for the nearly flat valence band (VB1) and conduction band (CB1). This diagram depicts the Chern number as a function of $\Delta$ and $\phi$ for specific values of $\alpha$. Also, we have shown the sinusoidal phase boundary separating the trivial and topological phase for the ordinary Haldane model of graphene, $\Delta = \pm 3 \sqrt{3}\sin{\phi}$ with a black curve \cite{PhysRevLett.61.2015}. 
It's essential to note that for intermediate values of $\alpha$ when the flat bands become dispersive, the topological phase depends on the NNN neighbor hopping strength $t_2$. We illustrate this by displaying the Chern number plots as a function of $\alpha$ for band VB1 with varying values of $t_2$. For instance, when $t_2 = 0.06t$ eV, we observe a Chern number phase transition from $|{C}| = 5$ to $|{C}| = 2$ at $\alpha = 0.559$, leading to a smaller $C = 2$ region in the Chern number phase diagram compared to $t_2 = 0.1t$ in Fig. \ref{fig:alpha_phi_phase}, irrespective of the value of $\phi$. As we increase $t_2$, the transition occurs at lower values of $\alpha$.      

\begin{comment}
\begin{figure}
    \centering
    \includegraphics[width=0.5\textwidth]{haldane_phs_n3_collage.pdf}
    \caption{The Haldane phase diagram shown for $VB1 (n = 3)$ for (A) $\alpha =1$ (Dice lattice), (B) $\alpha=0.6$ and (C) $\alpha = 0.2$ along with the sinusoidal Haldane phase boundary (black curve). The phase diagram for $CB1$ has the same feature with an opposite sign.}
    \label{fig:haldane_phase}
\end{figure}
\end{comment}

\begin{figure}
    \centering
    \includegraphics[width=0.5\textwidth]{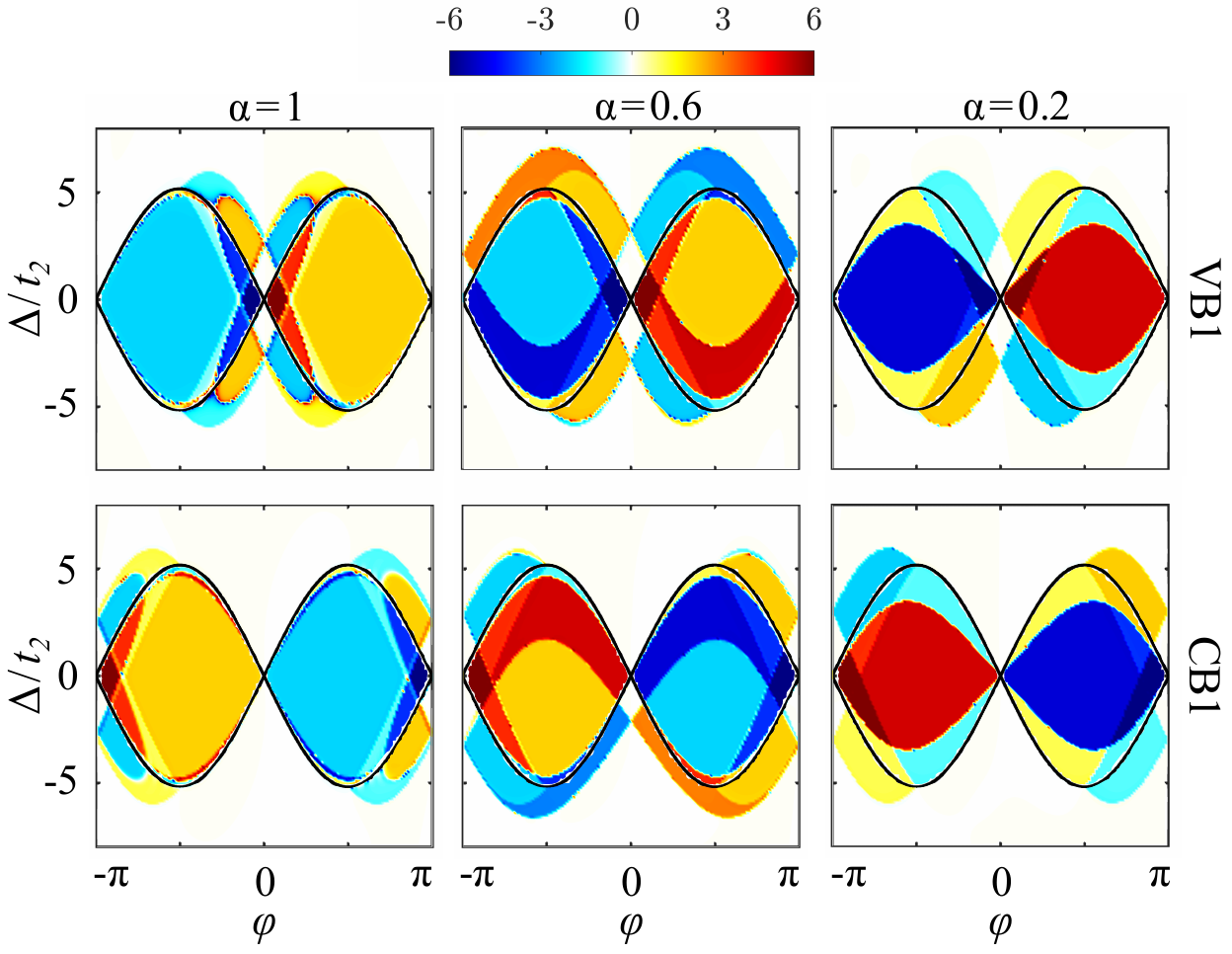}
    \caption{The Haldane phase diagram shown for $VB1 (n = 3)$ and $CB1 (n = 4)$ for different values of $\alpha$ along with the sinusoidal Haldane phase boundary (black curve).}
    \label{fig:haldane_phase}
\end{figure}

\subsection{Anomalous Hall Conductivity}
\begin{figure}
    \centering
    \includegraphics[width=0.32\textwidth]{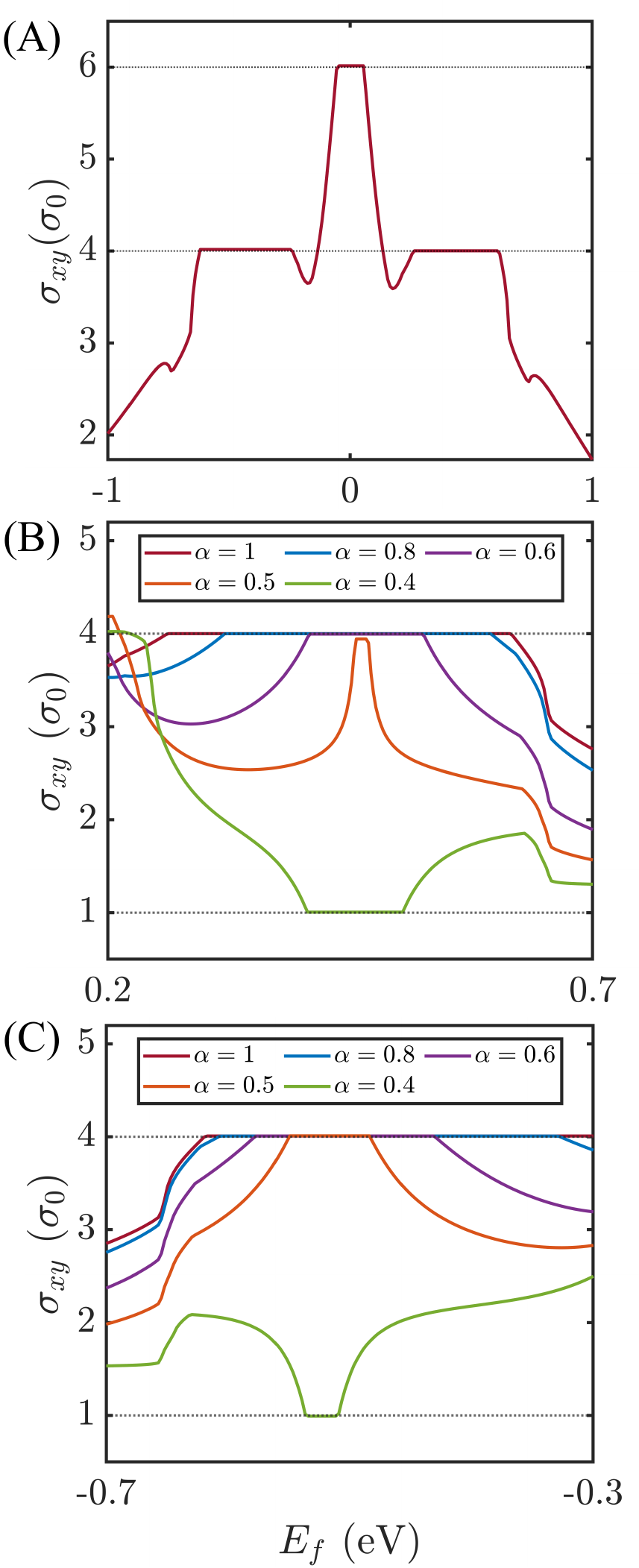}
    \caption{The hall conductivity depicted as a function of Fermi energy, $E_f$ (A) for $\alpha =1$, for several values of $\alpha$ for (B) Fermi energy near CB phase transition and (C) Fermi energy near VB phase transition, the other parameters taken as, $t_2=0.1t$, $\phi=\pi/2$, $\Delta=0$}
    \label{fig:qh_bilayer}
\end{figure}

In this section, we calculate the anomalous Hall conductivity for the bilayer cyclic $\alpha-\mathcal{T}_3$ lattice. This conductivity, a measure of how electrons respond to an external electric field, can provide insights into the topological properties of the material. We compute the Hall conductivity by numerically integrating the Berry curvature of all occupied electronic states over the entire Brillouin zone (BZ). This involves summing up contributions from all the bands in the material. \cite{RevModPhys.82.1959}, 
\begin{equation}
    \sigma_{xy} = \frac{\sigma_0}{2\pi} \sum_n \int \Omega_n (k_x,k_y) f(E_{k_x,k_y}^n) dk_x dk_y
\end{equation}
Where $\Omega_n$ is the berry curvature for $n^{th}$ band from Eq.(7), $f(E) = 1/[1+e^{(E-E_f)/{K_BT}}]$ is the Fermi-Dirac distribution function where $E_f$ and $T$ signifies Fermi energy and absolute temperature, $E_{k_x,k_y}^n$ is the energy eigenvalues for $n^{th}$ band and $\sigma_0 = e^2/h$. The anomalous hall conductivity arises due to the non-zero berry-curvature contributions of all occupied states. As the Fermi energy lies in a band gap, the Fermi-Dirac distribution function $f(E)$ at zero absolute temperature is one, and the total contribution of all occupied states comes from bands below the Fermi energy. Then the integration over the BZ gives the total Chern number of the bands below the Fermi energy and we get a plateau $\sigma_{xy} = |C_n|e^2/h$. The conductivity decays when the Fermi energy lies outside of the gap. The anomalous hall conductivity (AHC) for this system at $\alpha =1$ limit is plotted as a function of Fermi energy ($E_f$) in the unit of $\sigma_0$ as shown in Fig.\ref{fig:qh_bilayer}(A). From the AHC we noticed three different plateau regions with respect to the Fermi energies.

As we move the Fermi energy from zero energy to higher values to the gap between VB1 and VB2, the AHC is solely contributed by the VB3 ($C_n = 1$) and VB2 ($C_n = 3$), resulting into the $\sigma_{xy} = |4|e^2/h$ and the total Chern number $C_n = 4$ is just some of the two Chern numbers of VB3 and VB2. We see from the plot a plateau at $4e^2/h$ remains quantized within the gap. Similarly, we observe another plateau at $4e^2/h$ for the energy gap between CB2 and CB1 as the Chern number of VB1 and CB1 being $+2$ and $-2$, the total Chern number remains $4$. The system attains its maximum conductivity when the Fermi energy lies in the gap between VB1 and CB1. For the Fermi energy lies in this region, the total Chern number of all the bands below the Fermi energy is $6$ and we see a plateau at $6e^2/h$. The asymmetry in the plot arises due to the particle-hole asymmetry present in the Cyclic bilayer system. In order to visualize the nature of AHC plateaus after the phase transition, we have depicted the hall conductivity for several values of $\alpha$, for Fermi energies near the CB and VB phase transition point i.e. $E_f$ lying in the gap between CB1, CB2, and VB1, VB2 respectively. As we have discussed before (see Fig.\ref{fig:phase_transition_line_plt}), a topological phase transition accompanied by a band closure and opening is observed by changing hopping strength using  $\alpha$ for both CB and VB at $0.46$ eV and $-0.52$ eV respectively. And for the Fermi energy in any of the gaps mentioned above, we see a conductivity plateau at $4e^2/h$. As we lower the value of alpha, this plateau gradually diminishes as the gap becomes narrower and completely vanishes at the transition point when the band gap closes, and eventually, we see a new plateau arising at $e^2/h$ as the gap opens again.
The new plateau at $e^2/h$ arises because VB2 and CB2 become trivial bands after the transition which leads to the total Chern number of bands below the Fermi energy being $C_n = 1$ for both VB and CB. Fig.\ref{fig:qh_bilayer}(B) and \ref{fig:qh_bilayer}(C) show the conductivity for Fermi energies near the transition point for CB and VB respectively, for different $\alpha$ values.

\section{Conclusion}
We introduced the Haldane model for four non-equivalent vertically aligned commensurate stackings for bilayer $\alpha-\mathcal{T}_3$ lattice in the Dice lattice limit. We found that both Aligned and Cyclic stacking open up a band gap when the TRS is broken and the gap depends on both $t_2$ and inter-layer coupling. We investigated the topological properties of Cyclic stacking within the tight-binding formalism. The Cyclic stacking has all six bands topologically non-trivial including two nearly flat bands with a higher Chern number, $ C_n = \pm 3$. Further, we allowed the variation of $\alpha$ starting from the Dice lattice limit and observed two topological phase transitions in the valance band and the conduction band for two different values of $\alpha$. The phase transition occurs at the gap closing between VB1 and VB2 in the valance band and CB1 and CB2 in the conduction band leaving the Chern number of the third band unaltered. The partial flat bands acquire a Chern number as high as $C_n = \pm 5$ after the transition. The energy spectrum at $K$ and $K^\prime$ are independent of alpha and this is originally a feature of its single layer. 
We obtained the phase diagram in the parameter space $\alpha$ and $\phi$ for all six bands that exhibit multiple Chern phase regions. Higher Chern number arises for the low energy bands and a trivial region arises for higher energy bands. We also calculated the anomalous hall conductivity for the Dice lattice limit as well as for different $\alpha$ values for Fermi energy close to the transition point. The conductivity plateau shows a jump from $4e^2/h$ to $e^2/h$ which supports the phase transition from one Chern insulating phase to another. 

\begin{acknowledgments}
We acknowledge the support provided by the Kepler Computing facility, maintained by the Department of Physical Sciences, IISER Kolkata, for various computational needs. P.P. and S.G. acknowledge support from the Council of Scientific and Industrial Research (CSIR), India, for the doctoral fellowship. P.P. acknowledges Priyanka Sinha for useful discussions. B.L.C acknowledges the SERB with grant no. SRG/2022/001102 and ``IISER Kolkata Start-up-Grant" Ref. No. IISER-K/DoRD/SUG/BC/2021-22/376. P.P. thank Dr. Priyanka Sinha for the meaningful discussions.
\end{acknowledgments}
%\begin{subappendices}
%\appendix
\section*{APPENDIX}
\subsection*{Single layer \texorpdfstring{$\alpha-\mathcal{T}_3$}{} for \texorpdfstring{$\alpha \ne 1$}{}}
Here we discuss in more detail the effect of broken inversion symmetry in the presence of Haldane flux in single layer $\alpha-\mathcal{T}_3$ Hamiltonian (Eq.\eqref{eq:single_hamiltonian}) and the fate of topological bands and anomalous hall conductivity as we evolve the hopping strength between site $B$ and $C$ (Fig.\ref{fig: schematic}(a)) using the real parameter $\alpha$. 
\begin{figure*}
    \centering
    \includegraphics[scale=0.4]{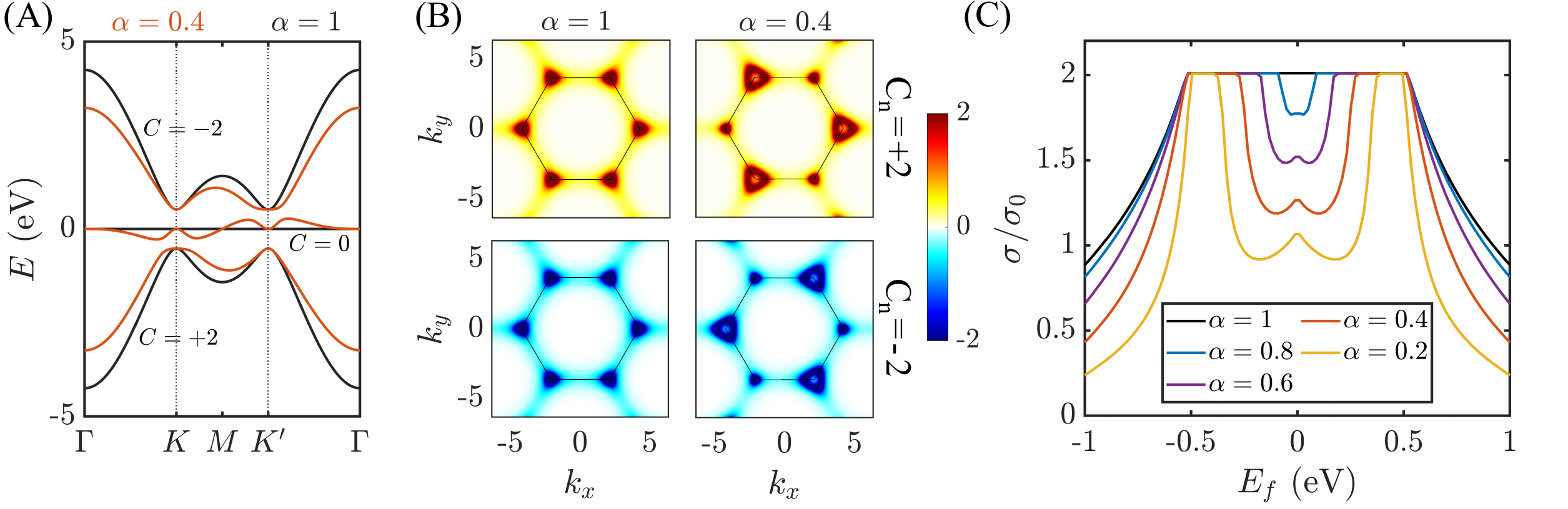}
    \caption{Single layer $\alpha-\mathcal{T}_3$ (A) spectrum for $\alpha = 1$ (black) and $\alpha = 0.4$ (red), (B) Berry curvatures for bands corresponding to the Chern number $+2$ (VB) and $-2$ (CB) are presented for $\alpha = 1$ (left) and $\alpha = 0.4$ (right). (C) Anomalous hall conductivity is shown for various values of $\alpha$, which are indicated in the inset. The values of $t_2$, $\phi$, $\Delta$ are taken as $0.1t$, $\pi/2$ and zero respectively.}
    \label{fig:single_layer}
\end{figure*}
Single layer $\alpha-\mathcal{T}_3$ lattice preserves the $W_0$ inversion symmetry in the dice lattice limit. For $\alpha$ values $\ne 0,1$ the inversion symmetry breaks. Introducing Haldane flux breaks time-reversal symmetry, Kramer's degeneracy is lifted and the energy becomes an odd function of $k$ as can be seen for the red curve ($\alpha = 0.4$) in the spectrum of Fig.\ref{fig:single_layer}(a). The Berry curvature plots for the VB and CB show different values at $K$ and $-K$ for $\alpha \ne 1$ showing the inversion symmetry breaking in the system. Though the middle band becomes dispersive and acquires a finite Berry curvature value, the total Berry curvature over BZ is zero and the band remains trivial in nature. Fig.\ref{fig:single_layer}(c) presents the hall conductivity as a function of Fermi energy for various values of alpha. $\alpha = 1$ corresponds to a completely flat middle band and this band has no contribution to the Berry curvature integral which leads to a hall conductivity plateau at $2\sigma_0$ when $E_f$ lies in the gap. When $\alpha \ne 1$ the dispersive middle band has a finite contribution to the Berry curvature integral and it reduces the overall contributions of all occupied states and we see a dip around zero energy. As the middle band becomes more dispersive with $\alpha$ values away from $1$, we see a broader and deeper dip in the hall conductivity. 

\bibliography{apssamp}% Produces the bibliography via BibTeX.

\begin{thebibliography}{54}%
\makeatletter
\providecommand \@ifxundefined [1]{%
 \@ifx{#1\undefined}
}%
\providecommand \@ifnum [1]{%
 \ifnum #1\expandafter \@firstoftwo
 \else \expandafter \@secondoftwo
 \fi
}%
\providecommand \@ifx [1]{%
 \ifx #1\expandafter \@firstoftwo
 \else \expandafter \@secondoftwo
 \fi
}%
\providecommand \natexlab [1]{#1}%
\providecommand \enquote  [1]{``#1''}%
\providecommand \bibnamefont  [1]{#1}%
\providecommand \bibfnamefont [1]{#1}%
\providecommand \citenamefont [1]{#1}%
\providecommand \href@noop [0]{\@secondoftwo}%
\providecommand \href [0]{\begingroup \@sanitize@url \@href}%
\providecommand \@href[1]{\@@startlink{#1}\@@href}%
\providecommand \@@href[1]{\endgroup#1\@@endlink}%
\providecommand \@sanitize@url [0]{\catcode `\\12\catcode `\$12\catcode
  `\&12\catcode `\#12\catcode `\^12\catcode `\_12\catcode `\%12\relax}%
\providecommand \@@startlink[1]{}%
\providecommand \@@endlink[0]{}%
\providecommand \url  [0]{\begingroup\@sanitize@url \@url }%
\providecommand \@url [1]{\endgroup\@href {#1}{\urlprefix }}%
\providecommand \urlprefix  [0]{URL }%
\providecommand \Eprint [0]{\href }%
\providecommand \doibase [0]{https://doi.org/}%
\providecommand \selectlanguage [0]{\@gobble}%
\providecommand \bibinfo  [0]{\@secondoftwo}%
\providecommand \bibfield  [0]{\@secondoftwo}%
\providecommand \translation [1]{[#1]}%
\providecommand \BibitemOpen [0]{}%
\providecommand \bibitemStop [0]{}%
\providecommand \bibitemNoStop [0]{.\EOS\space}%
\providecommand \EOS [0]{\spacefactor3000\relax}%
\providecommand \BibitemShut  [1]{\csname bibitem#1\endcsname}%
\let\auto@bib@innerbib\@empty
%</preamble>
\bibitem [{\citenamefont {Cao}\ \emph {et~al.}(2018)\citenamefont {Cao},
  \citenamefont {Fatemi}, \citenamefont {Fang}, \citenamefont {Watanabe},
  \citenamefont {Taniguchi}, \citenamefont {Kaxiras},\ and\ \citenamefont
  {Jarillo-Herrero}}]{Cao2018}%
  \BibitemOpen
  \bibfield  {author} {\bibinfo {author} {\bibfnamefont {Y.}~\bibnamefont
  {Cao}}, \bibinfo {author} {\bibfnamefont {V.}~\bibnamefont {Fatemi}},
  \bibinfo {author} {\bibfnamefont {S.}~\bibnamefont {Fang}}, \bibinfo {author}
  {\bibfnamefont {K.}~\bibnamefont {Watanabe}}, \bibinfo {author}
  {\bibfnamefont {T.}~\bibnamefont {Taniguchi}}, \bibinfo {author}
  {\bibfnamefont {E.}~\bibnamefont {Kaxiras}},\ and\ \bibinfo {author}
  {\bibfnamefont {P.}~\bibnamefont {Jarillo-Herrero}},\ }\bibfield  {title}
  {\bibinfo {title} {Unconventional superconductivity in magic-angle graphene
  superlattices},\ }\href {https://doi.org/10.1038/nature26160} {\bibfield
  {journal} {\bibinfo  {journal} {Nature}\ }\textbf {\bibinfo {volume} {556}},\
  \bibinfo {pages} {43} (\bibinfo {year} {2018})}\BibitemShut {NoStop}%
\bibitem [{\citenamefont {Liu}\ \emph {et~al.}(2021)\citenamefont {Liu},
  \citenamefont {Meng},\ and\ \citenamefont {Liu}}]{PhysRevMaterials.5.084203}%
  \BibitemOpen
  \bibfield  {author} {\bibinfo {author} {\bibfnamefont {H.}~\bibnamefont
  {Liu}}, \bibinfo {author} {\bibfnamefont {S.}~\bibnamefont {Meng}},\ and\
  \bibinfo {author} {\bibfnamefont {F.}~\bibnamefont {Liu}},\ }\bibfield
  {title} {\bibinfo {title} {Screening two-dimensional materials with
  topological flat bands},\ }\href
  {https://doi.org/10.1103/PhysRevMaterials.5.084203} {\bibfield  {journal}
  {\bibinfo  {journal} {Phys. Rev. Mater.}\ }\textbf {\bibinfo {volume} {5}},\
  \bibinfo {pages} {084203} (\bibinfo {year} {2021})}\BibitemShut {NoStop}%
\bibitem [{\citenamefont {Javvaji}\ \emph {et~al.}(2020)\citenamefont
  {Javvaji}, \citenamefont {Sun},\ and\ \citenamefont
  {Jung}}]{PhysRevB.101.125411}%
  \BibitemOpen
  \bibfield  {author} {\bibinfo {author} {\bibfnamefont {S.}~\bibnamefont
  {Javvaji}}, \bibinfo {author} {\bibfnamefont {J.-H.}\ \bibnamefont {Sun}},\
  and\ \bibinfo {author} {\bibfnamefont {J.}~\bibnamefont {Jung}},\ }\bibfield
  {title} {\bibinfo {title} {Topological flat bands without magic angles in
  massive twisted bilayer graphenes},\ }\href
  {https://doi.org/10.1103/PhysRevB.101.125411} {\bibfield  {journal} {\bibinfo
   {journal} {Phys. Rev. B}\ }\textbf {\bibinfo {volume} {101}},\ \bibinfo
  {pages} {125411} (\bibinfo {year} {2020})}\BibitemShut {NoStop}%
\bibitem [{\citenamefont {Yankowitz}\ \emph {et~al.}(2018)\citenamefont
  {Yankowitz}, \citenamefont {Jung}, \citenamefont {Laksono}, \citenamefont
  {Leconte}, \citenamefont {Chittari}, \citenamefont {Watanabe}, \citenamefont
  {Taniguchi}, \citenamefont {Adam}, \citenamefont {Graf},\ and\ \citenamefont
  {Dean}}]{Yankowitz2018}%
  \BibitemOpen
  \bibfield  {author} {\bibinfo {author} {\bibfnamefont {M.}~\bibnamefont
  {Yankowitz}}, \bibinfo {author} {\bibfnamefont {J.}~\bibnamefont {Jung}},
  \bibinfo {author} {\bibfnamefont {E.}~\bibnamefont {Laksono}}, \bibinfo
  {author} {\bibfnamefont {N.}~\bibnamefont {Leconte}}, \bibinfo {author}
  {\bibfnamefont {B.~L.}\ \bibnamefont {Chittari}}, \bibinfo {author}
  {\bibfnamefont {K.}~\bibnamefont {Watanabe}}, \bibinfo {author}
  {\bibfnamefont {T.}~\bibnamefont {Taniguchi}}, \bibinfo {author}
  {\bibfnamefont {S.}~\bibnamefont {Adam}}, \bibinfo {author} {\bibfnamefont
  {D.}~\bibnamefont {Graf}},\ and\ \bibinfo {author} {\bibfnamefont {C.~R.}\
  \bibnamefont {Dean}},\ }\bibfield  {title} {\bibinfo {title} {Dynamic
  band-structure tuning of graphene moir{\'e} superlattices with pressure},\
  }\href {https://doi.org/10.1038/s41586-018-0107-1} {\bibfield  {journal}
  {\bibinfo  {journal} {Nature}\ }\textbf {\bibinfo {volume} {557}},\ \bibinfo
  {pages} {404} (\bibinfo {year} {2018})}\BibitemShut {NoStop}%
\bibitem [{\citenamefont {Kim}\ \emph {et~al.}(2018)\citenamefont {Kim},
  \citenamefont {Leconte}, \citenamefont {Chittari}, \citenamefont {Watanabe},
  \citenamefont {Taniguchi}, \citenamefont {MacDonald}, \citenamefont {Jung},\
  and\ \citenamefont {Jung}}]{Kim.acs.nanolett.8b03423}%
  \BibitemOpen
  \bibfield  {author} {\bibinfo {author} {\bibfnamefont {H.}~\bibnamefont
  {Kim}}, \bibinfo {author} {\bibfnamefont {N.}~\bibnamefont {Leconte}},
  \bibinfo {author} {\bibfnamefont {B.~L.}\ \bibnamefont {Chittari}}, \bibinfo
  {author} {\bibfnamefont {K.}~\bibnamefont {Watanabe}}, \bibinfo {author}
  {\bibfnamefont {T.}~\bibnamefont {Taniguchi}}, \bibinfo {author}
  {\bibfnamefont {A.~H.}\ \bibnamefont {MacDonald}}, \bibinfo {author}
  {\bibfnamefont {J.}~\bibnamefont {Jung}},\ and\ \bibinfo {author}
  {\bibfnamefont {S.}~\bibnamefont {Jung}},\ }\bibfield  {title} {\bibinfo
  {title} {Accurate gap determination in monolayer and bilayer graphene/h-bn
  moiré superlattices},\ }\href {https://doi.org/10.1021/acs.nanolett.8b03423}
  {\bibfield  {journal} {\bibinfo  {journal} {Nano Letters}\ }\textbf {\bibinfo
  {volume} {18}},\ \bibinfo {pages} {7732} (\bibinfo {year} {2018})},\ \bibinfo
  {note} {pMID: 30457338},\ \Eprint
  {https://arxiv.org/abs/https://doi.org/10.1021/acs.nanolett.8b03423}
  {https://doi.org/10.1021/acs.nanolett.8b03423} \BibitemShut {NoStop}%
\bibitem [{\citenamefont {Chittari}\ \emph {et~al.}(2019)\citenamefont
  {Chittari}, \citenamefont {Chen}, \citenamefont {Zhang}, \citenamefont
  {Wang},\ and\ \citenamefont {Jung}}]{PhysRevLett.122.016401}%
  \BibitemOpen
  \bibfield  {author} {\bibinfo {author} {\bibfnamefont {B.~L.}\ \bibnamefont
  {Chittari}}, \bibinfo {author} {\bibfnamefont {G.}~\bibnamefont {Chen}},
  \bibinfo {author} {\bibfnamefont {Y.}~\bibnamefont {Zhang}}, \bibinfo
  {author} {\bibfnamefont {F.}~\bibnamefont {Wang}},\ and\ \bibinfo {author}
  {\bibfnamefont {J.}~\bibnamefont {Jung}},\ }\bibfield  {title} {\bibinfo
  {title} {Gate-tunable topological flat bands in trilayer graphene
  boron-nitride moir\'e superlattices},\ }\href
  {https://doi.org/10.1103/PhysRevLett.122.016401} {\bibfield  {journal}
  {\bibinfo  {journal} {Phys. Rev. Lett.}\ }\textbf {\bibinfo {volume} {122}},\
  \bibinfo {pages} {016401} (\bibinfo {year} {2019})}\BibitemShut {NoStop}%
\bibitem [{\citenamefont {Chen}\ \emph {et~al.}(2019)\citenamefont {Chen},
  \citenamefont {Jiang}, \citenamefont {Wu}, \citenamefont {Lyu}, \citenamefont
  {Li}, \citenamefont {Chittari}, \citenamefont {Watanabe}, \citenamefont
  {Taniguchi}, \citenamefont {Shi}, \citenamefont {Jung}, \citenamefont
  {Zhang},\ and\ \citenamefont {Wang}}]{Chen2019}%
  \BibitemOpen
  \bibfield  {author} {\bibinfo {author} {\bibfnamefont {G.}~\bibnamefont
  {Chen}}, \bibinfo {author} {\bibfnamefont {L.}~\bibnamefont {Jiang}},
  \bibinfo {author} {\bibfnamefont {S.}~\bibnamefont {Wu}}, \bibinfo {author}
  {\bibfnamefont {B.}~\bibnamefont {Lyu}}, \bibinfo {author} {\bibfnamefont
  {H.}~\bibnamefont {Li}}, \bibinfo {author} {\bibfnamefont {B.~L.}\
  \bibnamefont {Chittari}}, \bibinfo {author} {\bibfnamefont {K.}~\bibnamefont
  {Watanabe}}, \bibinfo {author} {\bibfnamefont {T.}~\bibnamefont {Taniguchi}},
  \bibinfo {author} {\bibfnamefont {Z.}~\bibnamefont {Shi}}, \bibinfo {author}
  {\bibfnamefont {J.}~\bibnamefont {Jung}}, \bibinfo {author} {\bibfnamefont
  {Y.}~\bibnamefont {Zhang}},\ and\ \bibinfo {author} {\bibfnamefont
  {F.}~\bibnamefont {Wang}},\ }\bibfield  {title} {\bibinfo {title} {Evidence
  of a gate-tunable mott insulator in a trilayer graphene moir{\'e}
  superlattice},\ }\href {https://doi.org/10.1038/s41567-018-0387-2} {\bibfield
   {journal} {\bibinfo  {journal} {Nature Physics}\ }\textbf {\bibinfo {volume}
  {15}},\ \bibinfo {pages} {237} (\bibinfo {year} {2019})}\BibitemShut
  {NoStop}%
\bibitem [{\citenamefont {Chebrolu}\ \emph {et~al.}(2019)\citenamefont
  {Chebrolu}, \citenamefont {Chittari},\ and\ \citenamefont
  {Jung}}]{PhysRevB.99.235417}%
  \BibitemOpen
  \bibfield  {author} {\bibinfo {author} {\bibfnamefont {N.~R.}\ \bibnamefont
  {Chebrolu}}, \bibinfo {author} {\bibfnamefont {B.~L.}\ \bibnamefont
  {Chittari}},\ and\ \bibinfo {author} {\bibfnamefont {J.}~\bibnamefont
  {Jung}},\ }\bibfield  {title} {\bibinfo {title} {Flat bands in twisted double
  bilayer graphene},\ }\href {https://doi.org/10.1103/PhysRevB.99.235417}
  {\bibfield  {journal} {\bibinfo  {journal} {Phys. Rev. B}\ }\textbf {\bibinfo
  {volume} {99}},\ \bibinfo {pages} {235417} (\bibinfo {year}
  {2019})}\BibitemShut {NoStop}%
\bibitem [{\citenamefont {Park}\ \emph {et~al.}(2020)\citenamefont {Park},
  \citenamefont {Chittari},\ and\ \citenamefont {Jung}}]{PhysRevB.102.035411}%
  \BibitemOpen
  \bibfield  {author} {\bibinfo {author} {\bibfnamefont {Y.}~\bibnamefont
  {Park}}, \bibinfo {author} {\bibfnamefont {B.~L.}\ \bibnamefont {Chittari}},\
  and\ \bibinfo {author} {\bibfnamefont {J.}~\bibnamefont {Jung}},\ }\bibfield
  {title} {\bibinfo {title} {Gate-tunable topological flat bands in twisted
  monolayer-bilayer graphene},\ }\href
  {https://doi.org/10.1103/PhysRevB.102.035411} {\bibfield  {journal} {\bibinfo
   {journal} {Phys. Rev. B}\ }\textbf {\bibinfo {volume} {102}},\ \bibinfo
  {pages} {035411} (\bibinfo {year} {2020})}\BibitemShut {NoStop}%
\bibitem [{\citenamefont {Sinha}\ \emph {et~al.}(2020)\citenamefont {Sinha},
  \citenamefont {Adak}, \citenamefont {Surya~Kanthi}, \citenamefont {Chittari},
  \citenamefont {Sangani}, \citenamefont {Watanabe}, \citenamefont {Taniguchi},
  \citenamefont {Jung},\ and\ \citenamefont {Deshmukh}}]{Sinha2020}%
  \BibitemOpen
  \bibfield  {author} {\bibinfo {author} {\bibfnamefont {S.}~\bibnamefont
  {Sinha}}, \bibinfo {author} {\bibfnamefont {P.~C.}\ \bibnamefont {Adak}},
  \bibinfo {author} {\bibfnamefont {R.~S.}\ \bibnamefont {Surya~Kanthi}},
  \bibinfo {author} {\bibfnamefont {B.~L.}\ \bibnamefont {Chittari}}, \bibinfo
  {author} {\bibfnamefont {L.~D.~V.}\ \bibnamefont {Sangani}}, \bibinfo
  {author} {\bibfnamefont {K.}~\bibnamefont {Watanabe}}, \bibinfo {author}
  {\bibfnamefont {T.}~\bibnamefont {Taniguchi}}, \bibinfo {author}
  {\bibfnamefont {J.}~\bibnamefont {Jung}},\ and\ \bibinfo {author}
  {\bibfnamefont {M.~M.}\ \bibnamefont {Deshmukh}},\ }\bibfield  {title}
  {\bibinfo {title} {Bulk valley transport and berry curvature spreading at the
  edge of flat bands},\ }\href {https://doi.org/10.1038/s41467-020-19284-w}
  {\bibfield  {journal} {\bibinfo  {journal} {Nature Communications}\ }\textbf
  {\bibinfo {volume} {11}},\ \bibinfo {pages} {5548} (\bibinfo {year}
  {2020})}\BibitemShut {NoStop}%
\bibitem [{\citenamefont {Shin}\ \emph
  {et~al.}(2021{\natexlab{a}})\citenamefont {Shin}, \citenamefont {Park},
  \citenamefont {Chittari}, \citenamefont {Sun},\ and\ \citenamefont
  {Jung}}]{PhysRevB.103.075423}%
  \BibitemOpen
  \bibfield  {author} {\bibinfo {author} {\bibfnamefont {J.}~\bibnamefont
  {Shin}}, \bibinfo {author} {\bibfnamefont {Y.}~\bibnamefont {Park}}, \bibinfo
  {author} {\bibfnamefont {B.~L.}\ \bibnamefont {Chittari}}, \bibinfo {author}
  {\bibfnamefont {J.-H.}\ \bibnamefont {Sun}},\ and\ \bibinfo {author}
  {\bibfnamefont {J.}~\bibnamefont {Jung}},\ }\bibfield  {title} {\bibinfo
  {title} {Electron-hole asymmetry and band gaps of commensurate double moire
  patterns in twisted bilayer graphene on hexagonal boron nitride},\ }\href
  {https://doi.org/10.1103/PhysRevB.103.075423} {\bibfield  {journal} {\bibinfo
   {journal} {Phys. Rev. B}\ }\textbf {\bibinfo {volume} {103}},\ \bibinfo
  {pages} {075423} (\bibinfo {year} {2021}{\natexlab{a}})}\BibitemShut
  {NoStop}%
\bibitem [{\citenamefont {Gonz\'alez}\ \emph {et~al.}(2021)\citenamefont
  {Gonz\'alez}, \citenamefont {Chittari}, \citenamefont {Park}, \citenamefont
  {Sun},\ and\ \citenamefont {Jung}}]{PhysRevB.103.165112}%
  \BibitemOpen
  \bibfield  {author} {\bibinfo {author} {\bibfnamefont {D.~A.~G.}\
  \bibnamefont {Gonz\'alez}}, \bibinfo {author} {\bibfnamefont {B.~L.}\
  \bibnamefont {Chittari}}, \bibinfo {author} {\bibfnamefont {Y.}~\bibnamefont
  {Park}}, \bibinfo {author} {\bibfnamefont {J.-H.}\ \bibnamefont {Sun}},\ and\
  \bibinfo {author} {\bibfnamefont {J.}~\bibnamefont {Jung}},\ }\bibfield
  {title} {\bibinfo {title} {Topological phases in $n$-layer abc graphene/boron
  nitride moir\'e superlattices},\ }\href
  {https://doi.org/10.1103/PhysRevB.103.165112} {\bibfield  {journal} {\bibinfo
   {journal} {Phys. Rev. B}\ }\textbf {\bibinfo {volume} {103}},\ \bibinfo
  {pages} {165112} (\bibinfo {year} {2021})}\BibitemShut {NoStop}%
\bibitem [{\citenamefont {Shin}\ \emph
  {et~al.}(2021{\natexlab{b}})\citenamefont {Shin}, \citenamefont {Chittari},\
  and\ \citenamefont {Jung}}]{PhysRevB.104.045413}%
  \BibitemOpen
  \bibfield  {author} {\bibinfo {author} {\bibfnamefont {J.}~\bibnamefont
  {Shin}}, \bibinfo {author} {\bibfnamefont {B.~L.}\ \bibnamefont {Chittari}},\
  and\ \bibinfo {author} {\bibfnamefont {J.}~\bibnamefont {Jung}},\ }\bibfield
  {title} {\bibinfo {title} {Stacking and gate-tunable topological flat bands,
  gaps, and anisotropic strip patterns in twisted trilayer graphene},\ }\href
  {https://doi.org/10.1103/PhysRevB.104.045413} {\bibfield  {journal} {\bibinfo
   {journal} {Phys. Rev. B}\ }\textbf {\bibinfo {volume} {104}},\ \bibinfo
  {pages} {045413} (\bibinfo {year} {2021}{\natexlab{b}})}\BibitemShut
  {NoStop}%
\bibitem [{\citenamefont {Shin}\ \emph {et~al.}(2022)\citenamefont {Shin},
  \citenamefont {Chittari}, \citenamefont {Jang}, \citenamefont {Min},\ and\
  \citenamefont {Jung}}]{PhysRevB.105.245124}%
  \BibitemOpen
  \bibfield  {author} {\bibinfo {author} {\bibfnamefont {J.}~\bibnamefont
  {Shin}}, \bibinfo {author} {\bibfnamefont {B.~L.}\ \bibnamefont {Chittari}},
  \bibinfo {author} {\bibfnamefont {Y.}~\bibnamefont {Jang}}, \bibinfo {author}
  {\bibfnamefont {H.}~\bibnamefont {Min}},\ and\ \bibinfo {author}
  {\bibfnamefont {J.}~\bibnamefont {Jung}},\ }\bibfield  {title} {\bibinfo
  {title} {Nearly flat bands in twisted triple bilayer graphene},\ }\href
  {https://doi.org/10.1103/PhysRevB.105.245124} {\bibfield  {journal} {\bibinfo
   {journal} {Phys. Rev. B}\ }\textbf {\bibinfo {volume} {105}},\ \bibinfo
  {pages} {245124} (\bibinfo {year} {2022})}\BibitemShut {NoStop}%
\bibitem [{\citenamefont {Chebrolu}\ and\ \citenamefont
  {Chittari}(2023)}]{CHEBROLU2023115526}%
  \BibitemOpen
  \bibfield  {author} {\bibinfo {author} {\bibfnamefont {N.~R.}\ \bibnamefont
  {Chebrolu}}\ and\ \bibinfo {author} {\bibfnamefont {B.~L.}\ \bibnamefont
  {Chittari}},\ }\bibfield  {title} {\bibinfo {title} {Analytical model of the
  energy spectrum and landau levels of a twisted double bilayer graphene},\
  }\href {https://doi.org/https://doi.org/10.1016/j.physe.2022.115526}
  {\bibfield  {journal} {\bibinfo  {journal} {Physica E: Low-dimensional
  Systems and Nanostructures}\ }\textbf {\bibinfo {volume} {146}},\ \bibinfo
  {pages} {115526} (\bibinfo {year} {2023})}\BibitemShut {NoStop}%
\bibitem [{\citenamefont {Park}\ \emph {et~al.}(2023)\citenamefont {Park},
  \citenamefont {Kim}, \citenamefont {Chittari},\ and\ \citenamefont
  {Jung}}]{PhysRevB.108.155406}%
  \BibitemOpen
  \bibfield  {author} {\bibinfo {author} {\bibfnamefont {Y.}~\bibnamefont
  {Park}}, \bibinfo {author} {\bibfnamefont {Y.}~\bibnamefont {Kim}}, \bibinfo
  {author} {\bibfnamefont {B.~L.}\ \bibnamefont {Chittari}},\ and\ \bibinfo
  {author} {\bibfnamefont {J.}~\bibnamefont {Jung}},\ }\bibfield  {title}
  {\bibinfo {title} {Topological flat bands in rhombohedral tetralayer and
  multilayer graphene on hexagonal boron nitride moir\'e superlattices},\
  }\href {https://doi.org/10.1103/PhysRevB.108.155406} {\bibfield  {journal}
  {\bibinfo  {journal} {Phys. Rev. B}\ }\textbf {\bibinfo {volume} {108}},\
  \bibinfo {pages} {155406} (\bibinfo {year} {2023})}\BibitemShut {NoStop}%
\bibitem [{\citenamefont {Miyahara}\ \emph {et~al.}(2007)\citenamefont
  {Miyahara}, \citenamefont {Kusuta},\ and\ \citenamefont
  {Furukawa}}]{MIYAHARA20071145}%
  \BibitemOpen
  \bibfield  {author} {\bibinfo {author} {\bibfnamefont {S.}~\bibnamefont
  {Miyahara}}, \bibinfo {author} {\bibfnamefont {S.}~\bibnamefont {Kusuta}},\
  and\ \bibinfo {author} {\bibfnamefont {N.}~\bibnamefont {Furukawa}},\
  }\bibfield  {title} {\bibinfo {title} {Bcs theory on a flat band lattice},\
  }\href {https://doi.org/https://doi.org/10.1016/j.physc.2007.03.393}
  {\bibfield  {journal} {\bibinfo  {journal} {Physica C: Superconductivity}\
  }\textbf {\bibinfo {volume} {460-462}},\ \bibinfo {pages} {1145} (\bibinfo
  {year} {2007})},\ \bibinfo {note} {proceedings of the 8th International
  Conference on Materials and Mechanisms of Superconductivity and High
  Temperature Superconductors}\BibitemShut {NoStop}%
\bibitem [{\citenamefont {J\'erome}\ \emph {et~al.}(1967)\citenamefont
  {J\'erome}, \citenamefont {Rice},\ and\ \citenamefont
  {Kohn}}]{PhysRev.158.462}%
  \BibitemOpen
  \bibfield  {author} {\bibinfo {author} {\bibfnamefont {D.}~\bibnamefont
  {J\'erome}}, \bibinfo {author} {\bibfnamefont {T.~M.}\ \bibnamefont {Rice}},\
  and\ \bibinfo {author} {\bibfnamefont {W.}~\bibnamefont {Kohn}},\ }\bibfield
  {title} {\bibinfo {title} {Excitonic insulator},\ }\href
  {https://doi.org/10.1103/PhysRev.158.462} {\bibfield  {journal} {\bibinfo
  {journal} {Phys. Rev.}\ }\textbf {\bibinfo {volume} {158}},\ \bibinfo {pages}
  {462} (\bibinfo {year} {1967})}\BibitemShut {NoStop}%
\bibitem [{\citenamefont {Sethi}\ \emph {et~al.}(2021)\citenamefont {Sethi},
  \citenamefont {Zhou}, \citenamefont {Zhu}, \citenamefont {Yang},\ and\
  \citenamefont {Liu}}]{PhysRevLett.126.196403}%
  \BibitemOpen
  \bibfield  {author} {\bibinfo {author} {\bibfnamefont {G.}~\bibnamefont
  {Sethi}}, \bibinfo {author} {\bibfnamefont {Y.}~\bibnamefont {Zhou}},
  \bibinfo {author} {\bibfnamefont {L.}~\bibnamefont {Zhu}}, \bibinfo {author}
  {\bibfnamefont {L.}~\bibnamefont {Yang}},\ and\ \bibinfo {author}
  {\bibfnamefont {F.}~\bibnamefont {Liu}},\ }\bibfield  {title} {\bibinfo
  {title} {Flat-band-enabled triplet excitonic insulator in a diatomic kagome
  lattice},\ }\href {https://doi.org/10.1103/PhysRevLett.126.196403} {\bibfield
   {journal} {\bibinfo  {journal} {Phys. Rev. Lett.}\ }\textbf {\bibinfo
  {volume} {126}},\ \bibinfo {pages} {196403} (\bibinfo {year}
  {2021})}\BibitemShut {NoStop}%
\bibitem [{\citenamefont {Tang}\ \emph {et~al.}(2011)\citenamefont {Tang},
  \citenamefont {Mei},\ and\ \citenamefont {Wen}}]{PhysRevLett.106.236802}%
  \BibitemOpen
  \bibfield  {author} {\bibinfo {author} {\bibfnamefont {E.}~\bibnamefont
  {Tang}}, \bibinfo {author} {\bibfnamefont {J.-W.}\ \bibnamefont {Mei}},\ and\
  \bibinfo {author} {\bibfnamefont {X.-G.}\ \bibnamefont {Wen}},\ }\bibfield
  {title} {\bibinfo {title} {High-temperature fractional quantum hall states},\
  }\href {https://doi.org/10.1103/PhysRevLett.106.236802} {\bibfield  {journal}
  {\bibinfo  {journal} {Phys. Rev. Lett.}\ }\textbf {\bibinfo {volume} {106}},\
  \bibinfo {pages} {236802} (\bibinfo {year} {2011})}\BibitemShut {NoStop}%
\bibitem [{\citenamefont {Neupert}\ \emph {et~al.}(2011)\citenamefont
  {Neupert}, \citenamefont {Santos}, \citenamefont {Chamon},\ and\
  \citenamefont {Mudry}}]{PhysRevLett.106.236804}%
  \BibitemOpen
  \bibfield  {author} {\bibinfo {author} {\bibfnamefont {T.}~\bibnamefont
  {Neupert}}, \bibinfo {author} {\bibfnamefont {L.}~\bibnamefont {Santos}},
  \bibinfo {author} {\bibfnamefont {C.}~\bibnamefont {Chamon}},\ and\ \bibinfo
  {author} {\bibfnamefont {C.}~\bibnamefont {Mudry}},\ }\bibfield  {title}
  {\bibinfo {title} {Fractional quantum hall states at zero magnetic field},\
  }\href {https://doi.org/10.1103/PhysRevLett.106.236804} {\bibfield  {journal}
  {\bibinfo  {journal} {Phys. Rev. Lett.}\ }\textbf {\bibinfo {volume} {106}},\
  \bibinfo {pages} {236804} (\bibinfo {year} {2011})}\BibitemShut {NoStop}%
\bibitem [{\citenamefont {Sun}\ \emph {et~al.}(2011)\citenamefont {Sun},
  \citenamefont {Gu}, \citenamefont {Katsura},\ and\ \citenamefont
  {Das~Sarma}}]{PhysRevLett.106.236803}%
  \BibitemOpen
  \bibfield  {author} {\bibinfo {author} {\bibfnamefont {K.}~\bibnamefont
  {Sun}}, \bibinfo {author} {\bibfnamefont {Z.}~\bibnamefont {Gu}}, \bibinfo
  {author} {\bibfnamefont {H.}~\bibnamefont {Katsura}},\ and\ \bibinfo {author}
  {\bibfnamefont {S.}~\bibnamefont {Das~Sarma}},\ }\bibfield  {title} {\bibinfo
  {title} {Nearly flatbands with nontrivial topology},\ }\href
  {https://doi.org/10.1103/PhysRevLett.106.236803} {\bibfield  {journal}
  {\bibinfo  {journal} {Phys. Rev. Lett.}\ }\textbf {\bibinfo {volume} {106}},\
  \bibinfo {pages} {236803} (\bibinfo {year} {2011})}\BibitemShut {NoStop}%
\bibitem [{\citenamefont {Stoner}(1938)}]{Stoner.rspa.1938.0066}%
  \BibitemOpen
  \bibfield  {author} {\bibinfo {author} {\bibfnamefont {E.~C.}\ \bibnamefont
  {Stoner}},\ }\bibfield  {title} {\bibinfo {title} {Collective electron
  ferromagnetism},\ }\href {https://doi.org/10.1098/rspa.1938.0066} {\bibfield
  {journal} {\bibinfo  {journal} {Proc. R. Soc. A.}\ }\textbf {\bibinfo
  {volume} {165}},\ \bibinfo {pages} {372} (\bibinfo {year}
  {1938})}\BibitemShut {NoStop}%
\bibitem [{\citenamefont {Jiang}\ \emph {et~al.}(2019)\citenamefont {Jiang},
  \citenamefont {Huang},\ and\ \citenamefont {Liu}}]{Jiang2019}%
  \BibitemOpen
  \bibfield  {author} {\bibinfo {author} {\bibfnamefont {W.}~\bibnamefont
  {Jiang}}, \bibinfo {author} {\bibfnamefont {H.}~\bibnamefont {Huang}},\ and\
  \bibinfo {author} {\bibfnamefont {F.}~\bibnamefont {Liu}},\ }\bibfield
  {title} {\bibinfo {title} {A lieb-like lattice in a covalent-organic
  framework and its stoner ferromagnetism},\ }\href
  {https://doi.org/10.1038/s41467-019-10094-3} {\bibfield  {journal} {\bibinfo
  {journal} {Nature Communications}\ }\textbf {\bibinfo {volume} {10}},\
  \bibinfo {pages} {2207} (\bibinfo {year} {2019})}\BibitemShut {NoStop}%
\bibitem [{\citenamefont {Zhou}\ \emph {et~al.}(2022)\citenamefont {Zhou},
  \citenamefont {Sethi}, \citenamefont {Liu}, \citenamefont {Wang},\ and\
  \citenamefont {Liu}}]{Zhou_2022}%
  \BibitemOpen
  \bibfield  {author} {\bibinfo {author} {\bibfnamefont {Y.}~\bibnamefont
  {Zhou}}, \bibinfo {author} {\bibfnamefont {G.}~\bibnamefont {Sethi}},
  \bibinfo {author} {\bibfnamefont {H.}~\bibnamefont {Liu}}, \bibinfo {author}
  {\bibfnamefont {Z.}~\bibnamefont {Wang}},\ and\ \bibinfo {author}
  {\bibfnamefont {F.}~\bibnamefont {Liu}},\ }\bibfield  {title} {\bibinfo
  {title} {Excited quantum anomalous and spin hall effect: dissociation of
  flat-bands-enabled excitonic insulator state},\ }\href
  {https://doi.org/10.1088/1361-6528/ac7a4b} {\bibfield  {journal} {\bibinfo
  {journal} {Nanotechnology}\ }\textbf {\bibinfo {volume} {33}},\ \bibinfo
  {pages} {415001} (\bibinfo {year} {2022})}\BibitemShut {NoStop}%
\bibitem [{\citenamefont {Ohgushi}\ \emph {et~al.}(2000)\citenamefont
  {Ohgushi}, \citenamefont {Murakami},\ and\ \citenamefont
  {Nagaosa}}]{PhysRevB.62.R6065}%
  \BibitemOpen
  \bibfield  {author} {\bibinfo {author} {\bibfnamefont {K.}~\bibnamefont
  {Ohgushi}}, \bibinfo {author} {\bibfnamefont {S.}~\bibnamefont {Murakami}},\
  and\ \bibinfo {author} {\bibfnamefont {N.}~\bibnamefont {Nagaosa}},\
  }\bibfield  {title} {\bibinfo {title} {Spin anisotropy and quantum hall
  effect in the kagom\'e lattice: Chiral spin state based on a ferromagnet},\
  }\href {https://doi.org/10.1103/PhysRevB.62.R6065} {\bibfield  {journal}
  {\bibinfo  {journal} {Phys. Rev. B}\ }\textbf {\bibinfo {volume} {62}},\
  \bibinfo {pages} {R6065} (\bibinfo {year} {2000})}\BibitemShut {NoStop}%
\bibitem [{\citenamefont {Guo}\ and\ \citenamefont
  {Franz}(2009)}]{PhysRevB.80.113102}%
  \BibitemOpen
  \bibfield  {author} {\bibinfo {author} {\bibfnamefont {H.-M.}\ \bibnamefont
  {Guo}}\ and\ \bibinfo {author} {\bibfnamefont {M.}~\bibnamefont {Franz}},\
  }\bibfield  {title} {\bibinfo {title} {Topological insulator on the kagome
  lattice},\ }\href {https://doi.org/10.1103/PhysRevB.80.113102} {\bibfield
  {journal} {\bibinfo  {journal} {Phys. Rev. B}\ }\textbf {\bibinfo {volume}
  {80}},\ \bibinfo {pages} {113102} (\bibinfo {year} {2009})}\BibitemShut
  {NoStop}%
\bibitem [{\citenamefont {Sutherland}(1986)}]{PhysRevB.34.5208}%
  \BibitemOpen
  \bibfield  {author} {\bibinfo {author} {\bibfnamefont {B.}~\bibnamefont
  {Sutherland}},\ }\bibfield  {title} {\bibinfo {title} {Localization of
  electronic wave functions due to local topology},\ }\href
  {https://doi.org/10.1103/PhysRevB.34.5208} {\bibfield  {journal} {\bibinfo
  {journal} {Phys. Rev. B}\ }\textbf {\bibinfo {volume} {34}},\ \bibinfo
  {pages} {5208} (\bibinfo {year} {1986})}\BibitemShut {NoStop}%
\bibitem [{\citenamefont {Vidal}\ \emph {et~al.}(1998)\citenamefont {Vidal},
  \citenamefont {Mosseri},\ and\ \citenamefont {Dou\ifmmode~\mbox{\c{c}}\else
  \c{c}\fi{}ot}}]{PhysRevLett.81.5888}%
  \BibitemOpen
  \bibfield  {author} {\bibinfo {author} {\bibfnamefont {J.}~\bibnamefont
  {Vidal}}, \bibinfo {author} {\bibfnamefont {R.}~\bibnamefont {Mosseri}},\
  and\ \bibinfo {author} {\bibfnamefont {B.}~\bibnamefont
  {Dou\ifmmode~\mbox{\c{c}}\else \c{c}\fi{}ot}},\ }\bibfield  {title} {\bibinfo
  {title} {Aharonov-bohm cages in two-dimensional structures},\ }\href
  {https://doi.org/10.1103/PhysRevLett.81.5888} {\bibfield  {journal} {\bibinfo
   {journal} {Phys. Rev. Lett.}\ }\textbf {\bibinfo {volume} {81}},\ \bibinfo
  {pages} {5888} (\bibinfo {year} {1998})}\BibitemShut {NoStop}%
\bibitem [{\citenamefont {Weeks}\ and\ \citenamefont
  {Franz}(2010)}]{PhysRevB.82.085310}%
  \BibitemOpen
  \bibfield  {author} {\bibinfo {author} {\bibfnamefont {C.}~\bibnamefont
  {Weeks}}\ and\ \bibinfo {author} {\bibfnamefont {M.}~\bibnamefont {Franz}},\
  }\bibfield  {title} {\bibinfo {title} {Topological insulators on the lieb and
  perovskite lattices},\ }\href {https://doi.org/10.1103/PhysRevB.82.085310}
  {\bibfield  {journal} {\bibinfo  {journal} {Phys. Rev. B}\ }\textbf {\bibinfo
  {volume} {82}},\ \bibinfo {pages} {085310} (\bibinfo {year}
  {2010})}\BibitemShut {NoStop}%
\bibitem [{\citenamefont {Goldman}\ \emph {et~al.}(2011)\citenamefont
  {Goldman}, \citenamefont {Urban},\ and\ \citenamefont
  {Bercioux}}]{PhysRevA.83.063601}%
  \BibitemOpen
  \bibfield  {author} {\bibinfo {author} {\bibfnamefont {N.}~\bibnamefont
  {Goldman}}, \bibinfo {author} {\bibfnamefont {D.~F.}\ \bibnamefont {Urban}},\
  and\ \bibinfo {author} {\bibfnamefont {D.}~\bibnamefont {Bercioux}},\
  }\bibfield  {title} {\bibinfo {title} {Topological phases for fermionic cold
  atoms on the lieb lattice},\ }\href
  {https://doi.org/10.1103/PhysRevA.83.063601} {\bibfield  {journal} {\bibinfo
  {journal} {Phys. Rev. A}\ }\textbf {\bibinfo {volume} {83}},\ \bibinfo
  {pages} {063601} (\bibinfo {year} {2011})}\BibitemShut {NoStop}%
\bibitem [{\citenamefont {and and}(2014)}]{Liu_2014}%
  \BibitemOpen
  \bibfield  {author} {\bibinfo {author} {\bibnamefont {and and}},\ }\bibfield
  {title} {\bibinfo {title} {Exotic electronic states in the world of flat
  bands: From theory to material},\ }\href
  {https://doi.org/10.1088/1674-1056/23/7/077308} {\bibfield  {journal}
  {\bibinfo  {journal} {Chinese Physics B}\ }\textbf {\bibinfo {volume} {23}},\
  \bibinfo {pages} {077308} (\bibinfo {year} {2014})}\BibitemShut {NoStop}%
\bibitem [{\citenamefont {Rhim}\ and\ \citenamefont
  {Yang}(2019)}]{PhysRevB.99.045107}%
  \BibitemOpen
  \bibfield  {author} {\bibinfo {author} {\bibfnamefont {J.-W.}\ \bibnamefont
  {Rhim}}\ and\ \bibinfo {author} {\bibfnamefont {B.-J.}\ \bibnamefont
  {Yang}},\ }\bibfield  {title} {\bibinfo {title} {Classification of flat bands
  according to the band-crossing singularity of bloch wave functions},\ }\href
  {https://doi.org/10.1103/PhysRevB.99.045107} {\bibfield  {journal} {\bibinfo
  {journal} {Phys. Rev. B}\ }\textbf {\bibinfo {volume} {99}},\ \bibinfo
  {pages} {045107} (\bibinfo {year} {2019})}\BibitemShut {NoStop}%
\bibitem [{\citenamefont {Bergman}\ \emph {et~al.}(2008)\citenamefont
  {Bergman}, \citenamefont {Wu},\ and\ \citenamefont
  {Balents}}]{PhysRevB.78.125104}%
  \BibitemOpen
  \bibfield  {author} {\bibinfo {author} {\bibfnamefont {D.~L.}\ \bibnamefont
  {Bergman}}, \bibinfo {author} {\bibfnamefont {C.}~\bibnamefont {Wu}},\ and\
  \bibinfo {author} {\bibfnamefont {L.}~\bibnamefont {Balents}},\ }\bibfield
  {title} {\bibinfo {title} {Band touching from real-space topology in
  frustrated hopping models},\ }\href
  {https://doi.org/10.1103/PhysRevB.78.125104} {\bibfield  {journal} {\bibinfo
  {journal} {Phys. Rev. B}\ }\textbf {\bibinfo {volume} {78}},\ \bibinfo
  {pages} {125104} (\bibinfo {year} {2008})}\BibitemShut {NoStop}%
\bibitem [{\citenamefont {Rizzi}\ \emph {et~al.}(2006)\citenamefont {Rizzi},
  \citenamefont {Cataudella},\ and\ \citenamefont
  {Fazio}}]{PhysRevB.73.144511}%
  \BibitemOpen
  \bibfield  {author} {\bibinfo {author} {\bibfnamefont {M.}~\bibnamefont
  {Rizzi}}, \bibinfo {author} {\bibfnamefont {V.}~\bibnamefont {Cataudella}},\
  and\ \bibinfo {author} {\bibfnamefont {R.}~\bibnamefont {Fazio}},\ }\bibfield
   {title} {\bibinfo {title} {Phase diagram of the bose-hubbard model with
  ${\mathcal{t}}_{3}$ symmetry},\ }\href
  {https://doi.org/10.1103/PhysRevB.73.144511} {\bibfield  {journal} {\bibinfo
  {journal} {Phys. Rev. B}\ }\textbf {\bibinfo {volume} {73}},\ \bibinfo
  {pages} {144511} (\bibinfo {year} {2006})}\BibitemShut {NoStop}%
\bibitem [{\citenamefont {Bercioux}\ \emph {et~al.}(2009)\citenamefont
  {Bercioux}, \citenamefont {Urban}, \citenamefont {Grabert},\ and\
  \citenamefont {H\"ausler}}]{PhysRevA.80.063603}%
  \BibitemOpen
  \bibfield  {author} {\bibinfo {author} {\bibfnamefont {D.}~\bibnamefont
  {Bercioux}}, \bibinfo {author} {\bibfnamefont {D.~F.}\ \bibnamefont {Urban}},
  \bibinfo {author} {\bibfnamefont {H.}~\bibnamefont {Grabert}},\ and\ \bibinfo
  {author} {\bibfnamefont {W.}~\bibnamefont {H\"ausler}},\ }\bibfield  {title}
  {\bibinfo {title} {Massless dirac-weyl fermions in a ${\mathcal{t}}_{3}$
  optical lattice},\ }\href {https://doi.org/10.1103/PhysRevA.80.063603}
  {\bibfield  {journal} {\bibinfo  {journal} {Phys. Rev. A}\ }\textbf {\bibinfo
  {volume} {80}},\ \bibinfo {pages} {063603} (\bibinfo {year}
  {2009})}\BibitemShut {NoStop}%
\bibitem [{\citenamefont {Wang}\ and\ \citenamefont
  {Ran}(2011)}]{PhysRevB.84.241103}%
  \BibitemOpen
  \bibfield  {author} {\bibinfo {author} {\bibfnamefont {F.}~\bibnamefont
  {Wang}}\ and\ \bibinfo {author} {\bibfnamefont {Y.}~\bibnamefont {Ran}},\
  }\bibfield  {title} {\bibinfo {title} {Nearly flat band with chern number
  $c=2$ on the dice lattice},\ }\href
  {https://doi.org/10.1103/PhysRevB.84.241103} {\bibfield  {journal} {\bibinfo
  {journal} {Phys. Rev. B}\ }\textbf {\bibinfo {volume} {84}},\ \bibinfo
  {pages} {241103} (\bibinfo {year} {2011})}\BibitemShut {NoStop}%
\bibitem [{\citenamefont {Illes}\ \emph {et~al.}(2015)\citenamefont {Illes},
  \citenamefont {Carbotte},\ and\ \citenamefont {Nicol}}]{PhysRevB.92.245410}%
  \BibitemOpen
  \bibfield  {author} {\bibinfo {author} {\bibfnamefont {E.}~\bibnamefont
  {Illes}}, \bibinfo {author} {\bibfnamefont {J.~P.}\ \bibnamefont
  {Carbotte}},\ and\ \bibinfo {author} {\bibfnamefont {E.~J.}\ \bibnamefont
  {Nicol}},\ }\bibfield  {title} {\bibinfo {title} {Hall quantization and
  optical conductivity evolution with variable berry phase in the
  $\ensuremath{\alpha}\text{\ensuremath{-}}{T}_{3}$ model},\ }\href
  {https://doi.org/10.1103/PhysRevB.92.245410} {\bibfield  {journal} {\bibinfo
  {journal} {Phys. Rev. B}\ }\textbf {\bibinfo {volume} {92}},\ \bibinfo
  {pages} {245410} (\bibinfo {year} {2015})}\BibitemShut {NoStop}%
\bibitem [{\citenamefont {Raoux}\ \emph {et~al.}(2014)\citenamefont {Raoux},
  \citenamefont {Morigi}, \citenamefont {Fuchs}, \citenamefont {Pi\'echon},\
  and\ \citenamefont {Montambaux}}]{PhysRevLett.112.026402}%
  \BibitemOpen
  \bibfield  {author} {\bibinfo {author} {\bibfnamefont {A.}~\bibnamefont
  {Raoux}}, \bibinfo {author} {\bibfnamefont {M.}~\bibnamefont {Morigi}},
  \bibinfo {author} {\bibfnamefont {J.-N.}\ \bibnamefont {Fuchs}}, \bibinfo
  {author} {\bibfnamefont {F.}~\bibnamefont {Pi\'echon}},\ and\ \bibinfo
  {author} {\bibfnamefont {G.}~\bibnamefont {Montambaux}},\ }\bibfield  {title}
  {\bibinfo {title} {From dia- to paramagnetic orbital susceptibility of
  massless fermions},\ }\href {https://doi.org/10.1103/PhysRevLett.112.026402}
  {\bibfield  {journal} {\bibinfo  {journal} {Phys. Rev. Lett.}\ }\textbf
  {\bibinfo {volume} {112}},\ \bibinfo {pages} {026402} (\bibinfo {year}
  {2014})}\BibitemShut {NoStop}%
\bibitem [{\citenamefont {Biswas}\ and\ \citenamefont
  {Ghosh}(2016)}]{biswas2016magnetotransport}%
  \BibitemOpen
  \bibfield  {author} {\bibinfo {author} {\bibfnamefont {T.}~\bibnamefont
  {Biswas}}\ and\ \bibinfo {author} {\bibfnamefont {T.~K.}\ \bibnamefont
  {Ghosh}},\ }\bibfield  {title} {\bibinfo {title} {Magnetotransport properties
  of the $\alpha$-t3 model},\ }\href@noop {} {\bibfield  {journal} {\bibinfo
  {journal} {Journal of Physics: Condensed Matter}\ }\textbf {\bibinfo {volume}
  {28}},\ \bibinfo {pages} {495302} (\bibinfo {year} {2016})}\BibitemShut
  {NoStop}%
\bibitem [{\citenamefont {Wang}\ \emph {et~al.}(2020)\citenamefont {Wang},
  \citenamefont {Liu}, \citenamefont {Wang},\ and\ \citenamefont
  {Liu}}]{PhysRevB.102.235414}%
  \BibitemOpen
  \bibfield  {author} {\bibinfo {author} {\bibfnamefont {J.~J.}\ \bibnamefont
  {Wang}}, \bibinfo {author} {\bibfnamefont {S.}~\bibnamefont {Liu}}, \bibinfo
  {author} {\bibfnamefont {J.}~\bibnamefont {Wang}},\ and\ \bibinfo {author}
  {\bibfnamefont {J.-F.}\ \bibnamefont {Liu}},\ }\bibfield  {title} {\bibinfo
  {title} {Integer quantum hall effect of the
  $\ensuremath{\alpha}\text{\ensuremath{-}}{T}_{3}$ model with a broken flat
  band},\ }\href {https://doi.org/10.1103/PhysRevB.102.235414} {\bibfield
  {journal} {\bibinfo  {journal} {Phys. Rev. B}\ }\textbf {\bibinfo {volume}
  {102}},\ \bibinfo {pages} {235414} (\bibinfo {year} {2020})}\BibitemShut
  {NoStop}%
\bibitem [{\citenamefont {Dey}\ and\ \citenamefont
  {Ghosh}(2018)}]{PhysRevB.98.075422}%
  \BibitemOpen
  \bibfield  {author} {\bibinfo {author} {\bibfnamefont {B.}~\bibnamefont
  {Dey}}\ and\ \bibinfo {author} {\bibfnamefont {T.~K.}\ \bibnamefont
  {Ghosh}},\ }\bibfield  {title} {\bibinfo {title} {Photoinduced valley and
  electron-hole symmetry breaking in $\ensuremath{\alpha}\ensuremath{-}{T}_{3}$
  lattice: The role of a variable berry phase},\ }\href
  {https://doi.org/10.1103/PhysRevB.98.075422} {\bibfield  {journal} {\bibinfo
  {journal} {Phys. Rev. B}\ }\textbf {\bibinfo {volume} {98}},\ \bibinfo
  {pages} {075422} (\bibinfo {year} {2018})}\BibitemShut {NoStop}%
\bibitem [{\citenamefont {Dey}\ and\ \citenamefont
  {Ghosh}(2019)}]{PhysRevB.99.205429}%
  \BibitemOpen
  \bibfield  {author} {\bibinfo {author} {\bibfnamefont {B.}~\bibnamefont
  {Dey}}\ and\ \bibinfo {author} {\bibfnamefont {T.~K.}\ \bibnamefont
  {Ghosh}},\ }\bibfield  {title} {\bibinfo {title} {Floquet topological phase
  transition in the $\ensuremath{\alpha}\text{\ensuremath{-}}{\mathcal{t}}_{3}$
  lattice},\ }\href {https://doi.org/10.1103/PhysRevB.99.205429} {\bibfield
  {journal} {\bibinfo  {journal} {Phys. Rev. B}\ }\textbf {\bibinfo {volume}
  {99}},\ \bibinfo {pages} {205429} (\bibinfo {year} {2019})}\BibitemShut
  {NoStop}%
\bibitem [{\citenamefont {Dey}\ \emph {et~al.}(2020)\citenamefont {Dey},
  \citenamefont {Kapri}, \citenamefont {Pal},\ and\ \citenamefont
  {Ghosh}}]{PhysRevB.101.235406}%
  \BibitemOpen
  \bibfield  {author} {\bibinfo {author} {\bibfnamefont {B.}~\bibnamefont
  {Dey}}, \bibinfo {author} {\bibfnamefont {P.}~\bibnamefont {Kapri}}, \bibinfo
  {author} {\bibfnamefont {O.}~\bibnamefont {Pal}},\ and\ \bibinfo {author}
  {\bibfnamefont {T.~K.}\ \bibnamefont {Ghosh}},\ }\bibfield  {title} {\bibinfo
  {title} {Unconventional phases in a haldane model of dice lattice},\ }\href
  {https://doi.org/10.1103/PhysRevB.101.235406} {\bibfield  {journal} {\bibinfo
   {journal} {Phys. Rev. B}\ }\textbf {\bibinfo {volume} {101}},\ \bibinfo
  {pages} {235406} (\bibinfo {year} {2020})}\BibitemShut {NoStop}%
\bibitem [{\citenamefont {Mondal}\ and\ \citenamefont
  {Basu}(2023)}]{PhysRevB.107.035421}%
  \BibitemOpen
  \bibfield  {author} {\bibinfo {author} {\bibfnamefont {S.}~\bibnamefont
  {Mondal}}\ and\ \bibinfo {author} {\bibfnamefont {S.}~\bibnamefont {Basu}},\
  }\bibfield  {title} {\bibinfo {title} {Topological features of the haldane
  model on a dice lattice: Flat-band effect on transport properties},\ }\href
  {https://doi.org/10.1103/PhysRevB.107.035421} {\bibfield  {journal} {\bibinfo
   {journal} {Phys. Rev. B}\ }\textbf {\bibinfo {volume} {107}},\ \bibinfo
  {pages} {035421} (\bibinfo {year} {2023})}\BibitemShut {NoStop}%
\bibitem [{\citenamefont {Sukhachov}\ \emph {et~al.}(2023)\citenamefont
  {Sukhachov}, \citenamefont {Oriekhov},\ and\ \citenamefont
  {Gorbar}}]{PhysRevB.108.075166}%
  \BibitemOpen
  \bibfield  {author} {\bibinfo {author} {\bibfnamefont {P.~O.}\ \bibnamefont
  {Sukhachov}}, \bibinfo {author} {\bibfnamefont {D.~O.}\ \bibnamefont
  {Oriekhov}},\ and\ \bibinfo {author} {\bibfnamefont {E.~V.}\ \bibnamefont
  {Gorbar}},\ }\bibfield  {title} {\bibinfo {title} {Stackings and effective
  models of bilayer dice lattices},\ }\href
  {https://doi.org/10.1103/PhysRevB.108.075166} {\bibfield  {journal} {\bibinfo
   {journal} {Phys. Rev. B}\ }\textbf {\bibinfo {volume} {108}},\ \bibinfo
  {pages} {075166} (\bibinfo {year} {2023})}\BibitemShut {NoStop}%
\bibitem [{\citenamefont {Korshunov}(2001)}]{PhysRevB.63.134503}%
  \BibitemOpen
  \bibfield  {author} {\bibinfo {author} {\bibfnamefont {S.~E.}\ \bibnamefont
  {Korshunov}},\ }\bibfield  {title} {\bibinfo {title} {Vortex ordering in
  fully frustrated superconducting systems with a dice lattice},\ }\href
  {https://doi.org/10.1103/PhysRevB.63.134503} {\bibfield  {journal} {\bibinfo
  {journal} {Phys. Rev. B}\ }\textbf {\bibinfo {volume} {63}},\ \bibinfo
  {pages} {134503} (\bibinfo {year} {2001})}\BibitemShut {NoStop}%
\bibitem [{\citenamefont {Urban}\ \emph {et~al.}(2011)\citenamefont {Urban},
  \citenamefont {Bercioux}, \citenamefont {Wimmer},\ and\ \citenamefont
  {H\"ausler}}]{PhysRevB.84.115136}%
  \BibitemOpen
  \bibfield  {author} {\bibinfo {author} {\bibfnamefont {D.~F.}\ \bibnamefont
  {Urban}}, \bibinfo {author} {\bibfnamefont {D.}~\bibnamefont {Bercioux}},
  \bibinfo {author} {\bibfnamefont {M.}~\bibnamefont {Wimmer}},\ and\ \bibinfo
  {author} {\bibfnamefont {W.}~\bibnamefont {H\"ausler}},\ }\bibfield  {title}
  {\bibinfo {title} {Barrier transmission of dirac-like pseudospin-one
  particles},\ }\href {https://doi.org/10.1103/PhysRevB.84.115136} {\bibfield
  {journal} {\bibinfo  {journal} {Phys. Rev. B}\ }\textbf {\bibinfo {volume}
  {84}},\ \bibinfo {pages} {115136} (\bibinfo {year} {2011})}\BibitemShut
  {NoStop}%
\bibitem [{\citenamefont {Malcolm}\ and\ \citenamefont
  {Nicol}(2016)}]{PhysRevB.93.165433}%
  \BibitemOpen
  \bibfield  {author} {\bibinfo {author} {\bibfnamefont {J.~D.}\ \bibnamefont
  {Malcolm}}\ and\ \bibinfo {author} {\bibfnamefont {E.~J.}\ \bibnamefont
  {Nicol}},\ }\bibfield  {title} {\bibinfo {title} {Frequency-dependent
  polarizability, plasmons, and screening in the two-dimensional pseudospin-1
  dice lattice},\ }\href {https://doi.org/10.1103/PhysRevB.93.165433}
  {\bibfield  {journal} {\bibinfo  {journal} {Phys. Rev. B}\ }\textbf {\bibinfo
  {volume} {93}},\ \bibinfo {pages} {165433} (\bibinfo {year}
  {2016})}\BibitemShut {NoStop}%
\bibitem [{\citenamefont {Vigh}\ \emph {et~al.}(2013)\citenamefont {Vigh},
  \citenamefont {Oroszl\'any}, \citenamefont {Vajna}, \citenamefont {San-Jose},
  \citenamefont {D\'avid}, \citenamefont {Cserti},\ and\ \citenamefont
  {D\'ora}}]{PhysRevB.88.161413}%
  \BibitemOpen
  \bibfield  {author} {\bibinfo {author} {\bibfnamefont {M.}~\bibnamefont
  {Vigh}}, \bibinfo {author} {\bibfnamefont {L.}~\bibnamefont {Oroszl\'any}},
  \bibinfo {author} {\bibfnamefont {S.}~\bibnamefont {Vajna}}, \bibinfo
  {author} {\bibfnamefont {P.}~\bibnamefont {San-Jose}}, \bibinfo {author}
  {\bibfnamefont {G.}~\bibnamefont {D\'avid}}, \bibinfo {author} {\bibfnamefont
  {J.}~\bibnamefont {Cserti}},\ and\ \bibinfo {author} {\bibfnamefont
  {B.}~\bibnamefont {D\'ora}},\ }\bibfield  {title} {\bibinfo {title}
  {Diverging dc conductivity due to a flat band in a disordered system of
  pseudospin-1 dirac-weyl fermions},\ }\href
  {https://doi.org/10.1103/PhysRevB.88.161413} {\bibfield  {journal} {\bibinfo
  {journal} {Phys. Rev. B}\ }\textbf {\bibinfo {volume} {88}},\ \bibinfo
  {pages} {161413} (\bibinfo {year} {2013})}\BibitemShut {NoStop}%
\bibitem [{\citenamefont {Xiao}\ \emph {et~al.}(2010)\citenamefont {Xiao},
  \citenamefont {Chang},\ and\ \citenamefont {Niu}}]{RevModPhys.82.1959}%
  \BibitemOpen
  \bibfield  {author} {\bibinfo {author} {\bibfnamefont {D.}~\bibnamefont
  {Xiao}}, \bibinfo {author} {\bibfnamefont {M.-C.}\ \bibnamefont {Chang}},\
  and\ \bibinfo {author} {\bibfnamefont {Q.}~\bibnamefont {Niu}},\ }\bibfield
  {title} {\bibinfo {title} {Berry phase effects on electronic properties},\
  }\href {https://doi.org/10.1103/RevModPhys.82.1959} {\bibfield  {journal}
  {\bibinfo  {journal} {Rev. Mod. Phys.}\ }\textbf {\bibinfo {volume} {82}},\
  \bibinfo {pages} {1959} (\bibinfo {year} {2010})}\BibitemShut {NoStop}%
\bibitem [{\citenamefont {Thouless}\ \emph {et~al.}(1982)\citenamefont
  {Thouless}, \citenamefont {Kohmoto}, \citenamefont {Nightingale},\ and\
  \citenamefont {den Nijs}}]{PhysRevLett.49.405}%
  \BibitemOpen
  \bibfield  {author} {\bibinfo {author} {\bibfnamefont {D.~J.}\ \bibnamefont
  {Thouless}}, \bibinfo {author} {\bibfnamefont {M.}~\bibnamefont {Kohmoto}},
  \bibinfo {author} {\bibfnamefont {M.~P.}\ \bibnamefont {Nightingale}},\ and\
  \bibinfo {author} {\bibfnamefont {M.}~\bibnamefont {den Nijs}},\ }\bibfield
  {title} {\bibinfo {title} {Quantized hall conductance in a two-dimensional
  periodic potential},\ }\href {https://doi.org/10.1103/PhysRevLett.49.405}
  {\bibfield  {journal} {\bibinfo  {journal} {Phys. Rev. Lett.}\ }\textbf
  {\bibinfo {volume} {49}},\ \bibinfo {pages} {405} (\bibinfo {year}
  {1982})}\BibitemShut {NoStop}%
\bibitem [{\citenamefont {Avron}\ \emph {et~al.}(1983)\citenamefont {Avron},
  \citenamefont {Seiler},\ and\ \citenamefont {Simon}}]{PhysRevLett.51.51}%
  \BibitemOpen
  \bibfield  {author} {\bibinfo {author} {\bibfnamefont {J.~E.}\ \bibnamefont
  {Avron}}, \bibinfo {author} {\bibfnamefont {R.}~\bibnamefont {Seiler}},\ and\
  \bibinfo {author} {\bibfnamefont {B.}~\bibnamefont {Simon}},\ }\bibfield
  {title} {\bibinfo {title} {Homotopy and quantization in condensed matter
  physics},\ }\href {https://doi.org/10.1103/PhysRevLett.51.51} {\bibfield
  {journal} {\bibinfo  {journal} {Phys. Rev. Lett.}\ }\textbf {\bibinfo
  {volume} {51}},\ \bibinfo {pages} {51} (\bibinfo {year} {1983})}\BibitemShut
  {NoStop}%
\bibitem [{\citenamefont {Haldane}(1988)}]{PhysRevLett.61.2015}%
  \BibitemOpen
  \bibfield  {author} {\bibinfo {author} {\bibfnamefont {F.~D.~M.}\
  \bibnamefont {Haldane}},\ }\bibfield  {title} {\bibinfo {title} {Model for a
  quantum hall effect without landau levels: Condensed-matter realization of
  the "parity anomaly"},\ }\href {https://doi.org/10.1103/PhysRevLett.61.2015}
  {\bibfield  {journal} {\bibinfo  {journal} {Phys. Rev. Lett.}\ }\textbf
  {\bibinfo {volume} {61}},\ \bibinfo {pages} {2015} (\bibinfo {year}
  {1988})}\BibitemShut {NoStop}%
\end{thebibliography}%

\end{document}